\documentclass[10pt,journal]{IEEEtran}
\usepackage{amsmath}
\usepackage{amssymb}
\usepackage{amsfonts}

\newcommand{\Rmnum}[1]{\expandafter\@slowromancap\romannumeral #1@}
\usepackage{graphicx}
\usepackage{placeins}
\usepackage{needspace}
\usepackage{epsfig}
\usepackage[sort]{cite}
\usepackage{xcolor}
\usepackage{enumerate}
\usepackage{extarrows}
\usepackage{algorithmic}
\usepackage{stfloats}
\usepackage{subfigure}
\usepackage{bbm}   
\usepackage[lined,ruled,linesnumbered]{algorithm2e}
\usepackage{tensor}
\usepackage[colorlinks,linkcolor=blue,anchorcolor=blue,citecolor=blue,urlcolor=black]{hyperref}
\DeclareRobustCommand{\revision}[1]{{\color{black}#1}}
\begin{document}
	\title{Reliable Near-Field Source Enumeration and Positioning Informed by Two-Stage MUSIC}
	
\author{Jiaying Li, Haifeng Wen, Changsheng You, Yuanwei Liu, and Hong Xing%
\thanks{J. Li, H. Wen, and H. Xing are with The Hong Kong University of Science and Technology (Guangzhou), Guangzhou, 511453, China; H. Xing is also affiliated with The Hong Kong University of Science and Technology, HK SAR (e-mails: jli989@connect.hkust-gz.edu.cn, hwen904@connect.hkust-gz.edu.cn, hongxing@ust.hk).}%
\thanks{C. You is with Southern University of Science and Technology, Shenzhen, China (e-mail: youcs@sustech.edu.cn).}%
\thanks{Y. Liu is with The University of Hong Kong, HK SAR (e-mail: yuanwei@hku.hk).}%
}
	\maketitle

	\begin{abstract}
		Accurate and reliable near-field multi-source localization is essential for location-aware processing in future wireless systems. However, coherent multipath propagation complicates the direct recovery of line-of-sight (LoS) source positions, while existing estimation methods generally lack explicit control of the risk of source underestimation and positioning uncertainty. In this paper, we investigate reliable source enumeration and positioning in near-field coherent multipath environments. Specifically, we develop a conformal risk control (CRC) based source enumeration scheme that calibrates the penalty coefficient of an eigenvalue-based criterion to control the expected miss-detection ratio at a prescribed level. We further propose \emph{MUSIC-Net}, a model-based deep learning framework informed by two-stage MUltiple SIgnal Classification (MUSIC), which learns the LoS-related subspace and surrogate distances through subspace-distance-based training, thereby enabling direct source positioning without estimating non-line-of-sight (NLoS) path parameters or performing subsequent path association. To quantify the uncertainty of the estimated positions, we develop a split conformal prediction (SCP) based calibration method that constructs angle–range predicted regions with finite-sample simultaneous coverage guarantees. Numerical results show that the proposed framework effectively controls the risk of source underestimation, while providing improved positioning accuracy with tighter predicted regions than benchmarks given the same target coverage levels.
	\end{abstract}
	
	\IEEEpeerreviewmaketitle
	\newtheorem{definition}{\underline{Definition}}[section]
	\newtheorem{fact}{Fact}
	\newtheorem{assumption}{Assumption}
	\newtheorem{theorem}{\underline{Theorem}}[section]
	\newtheorem{lemma}{\underline{Lemma}}[section]
	\newtheorem{proposition}{\underline{Proposition}}[section]
	\newtheorem{corollary}[proposition]{\underline{Corollary}}
	\newtheorem{example}{\underline{Example}}[section]
	\newtheorem{remark}{\underline{Remark}}[section]
	\newcommand{\mv}[1]{\mbox{\boldmath{$ #1 $}}}
	\newcommand{\mb}[1]{\mathbb{#1}}
	\newcommand{\Myfrac}[2]{\ensuremath{#1\mathord{\left/\right.\kern-\nulldelimiterspace}#2}}
	\newcommand\Perms[2]{\tensor[^{#2}]P{_{#1}}}

	\section{Introduction}
		
		
	
	Future wireless systems are expected to increasingly exploit extremely large-scale antenna arrays~\cite{wang2024tutorial} and high-frequency spectrum resources, such as millimeter-wave~\cite{wang2018millimeter} and sub-terahertz bands~\cite{Thomas_2025_THz}, to support high-capacity and highly directional wireless communications~\cite{liu2023near}. The combination of large array apertures and short wavelengths gives rise to two important considerations. First, highly focused near-field transmission becomes increasingly dependent on accurate source location information, which is essential for beam focusing and spatial processing~\cite{zhang20236g}. Second, the enlarged Fraunhofer distance places an increasing number of sources in the radiative near-field~\cite{zhang20236g}, where spherical-wave propagation becomes non-negligible and the conventional planar-wave assumption is no longer accurate~\cite{liu2023near}. Unlike far-field propagation, where the array response mainly depends on the direction of arrival (DoA) and is insensitive to the source range,  near-field spherical waveforms exhibit angle- and range-dependent spatial characteristics, thereby entailing the joint recovery of source direction and distance~\cite{yang2025near}. These unique propagation characteristics have stimulated growing interest in high-resolution near-field multi-source localization for future wireless systems~\cite{lei2025near_field}.
	
	Near-field multi-source localization generally involves two fundamental tasks, namely source enumeration and spatial parameter estimation~\cite{yang2025beamforming}. The former determines the number of sources present in the received observations, whereas the latter estimates their angle--range parameters or positions. In practical propagation environments, however, both tasks are subject to uncertainty. Source enumeration may underestimate or overestimate the number of sources under adverse signal conditions, while spatial parameter estimation is challenged by coherent multipath propagation and uncertainty in the resulting position estimates. Since the estimated source number determines the number of sources to be localized and the signal--noise subspace partition, errors in source enumeration can directly affect downstream positioning. These considerations motivate reliable processing across both stages of near-field multi-source localization.
	
	\subsection{Related Work}

	Traditional source enumeration methods mainly include information-theoretic criterion (ITC)-based methods~\cite{stoica2004model}, eigenvalue-thresholding methods~\cite{chen1991detection,kritchman2009non}, and compressive sensing (CS)-based methods~\cite{hu2014near,hu2014near1}. Compared with eigenvalue-thresholding methods, which rely on prescribed or adaptively designed decision thresholds, and CS-based methods, which generally require spatial dictionary construction and sparse recovery, leading to increased computational complexity and sensitivity to model mismatch, ITC-based methods provide a simple and interpretable model-order selection framework by balancing data fitting and model complexity. Consequently, ITC-based methods, such as the Akaike information criterion (AIC)~\cite{akaike1998information}, generalized information criterion (GIC)~\cite{bhansali1977some}, and minimum description length (MDL) criterion~\cite{rissanen1978modeling}, have been widely adopted for source enumeration. More recently, due to their powerful feature extraction capability, data-driven methods based on deep learning (DL) have also been introduced into source enumeration to counter challenging signal conditions such as low signal-to-noise ratio (SNR)  and model mismatch. For example, reference~\cite{barthelme2021machine} treated source-number estimation as a multiclass classification problem and employed the cross-entropy loss for network training. Reference~\cite{gast2025dcd} designed an eigenspectrum regularization loss to enhance the separability between signal and noise eigenvalues for threshold-based source enumeration. Reference~\cite{Gast2026NFsubspace} further proposed model-order-aware loss functions tailored to ITC-based methods such as AIC and MDL.
	
	Turning to spatial parameter estimation, near-field localization has been extensively investigated using beamforming~\cite{yang2025beamforming}, maximum-likelihood (ML) estimation~\cite{cheng2022efficient}, CS~\cite{rinchi2022cs,xu2025userlocalizationchannelestimation}, and subspace-based methods~\cite{challa1995high,yuen2002performance1,zuo2018new,huang2002near,zhi2007near,zhang2018rdmusic,gui_2020_cumulant,qu2024two}. Among them, two-dimension (2D) MUltiple SIgnal Classification (MUSIC)~\cite{huang2002near} and its variants~\cite{zhi2007near,zhang2018rdmusic,gui_2020_cumulant,qu2024two} are attractive due to their high resolution and physically interpretable spatial spectra. However, the joint angle--range search of 2D MUSIC can incur considerable computational complexity as the search resolution increases~\cite{zhang2018rdmusic}. To alleviate this burden, several multi-stage MUSIC variants~\cite{zhi2007near,zhang2018rdmusic,gui_2020_cumulant,qu2024two} decouple angle and range estimation into sequential one-dimension (1D) searches. In particular, Two-Stage MUSIC~\cite{qu2024two} introduced a surrogate distance for angular search and subsequently searched distances conditioned on the recovered directions, thereby avoiding the restrictive antenna-spacing conditions required by some earlier reduced-dimension methods~\cite{zhi2007near,zhang2018rdmusic,gui_2020_cumulant}.
	
	DL has also been incorporated into near-field localization to improve position estimation performance. Existing end-to-end (E2E) approaches directly learn the mapping from received observations to localization outputs using architectures such as convolutional neural networks (CNNs)~\cite{su2021cnn_near}. Although offering considerable modeling flexibility, such E2E approaches often underutilize known array and propagation structures, while requiring highly parameterized architectures and providing limited physical interpretability. By contrast, model-based DL embeds trainable deep neural networks (DNNs) into physically structured signal processing pipelines and learns intermediate representations for subsequent model-based inference~\cite{shlezinger2023model}. Following this principle, recent DL-aided subspace methods learn covariance representations tailored to subsequent subspace-based localization algorithms~\cite{Gast2026NFsubspace,gast2025dcd,zhu2025transformer}. Representative examples include NF-SubspaceNet~\cite{Gast2026NFsubspace}, which learned a surrogate covariance matrix tailored to 2D MUSIC-based localization, and DCD-MUSIC~\cite{gast2025dcd}, which learned stage-specific surrogate covariance matrices for cascaded angle and distance estimation.

	Conformal prediction (CP)~\cite{angelopoulos2023conformal} provides a model-free and distribution-free uncertainty quantification (UQ) framework for constructing prediction sets with statistical coverage guarantees, while its extension, conformal risk control (CRC)~\cite{angelopoulos_conformal_2024}, enables calibration under general task-specific risk functions and explicit control of asymmetric risks such as source underestimation. CP and CRC have recently found applications in intelligent wireless systems, including channel prediction~\cite{cohen2023calibrating}, wireless resource allocation~\cite{binucci2025conformal}, beam selection~\cite{deng2026scan}, and sensing~\cite{khurjekar2023doa}. For instance, reference~\cite{cohen2023calibrating} employed CP to construct channel prediction intervals with prescribed coverage. Reference~\cite{khurjekar2023doa} applied CP to calibrate prediction intervals for single-source DoA estimation in far-field scenarios. Reference~\cite{binucci2025conformal} incorporated CRC to control the false negative rate in resource allocation, while reference~\cite{deng2026scan} used CRC to control the probability of excluding near-optimal beams from the candidate beam set.
	
	\subsection{Motivation and Contributions}
	Based on the preceding review of related work, reliable near-field multi-source localization in coherent multipath environments faces the following three key challenges:
	
	First, when the operating environment changes, for example, due to low SNR, limited snapshots, or variations in propagation and source distributions, existing source enumeration methods may suffer performance degradation, resulting in both source underestimation and overestimation. In particular, standard criteria such as AIC~\cite{akaike1998information} and MDL~\cite{rissanen1978modeling} do not explicitly control the expected miss-detection rate (MDR) at a prescribed level. Moreover, conformal calibration remains largely unexplored for reliable source enumeration in near-field multi-source localization.
	
	Second, coherent multipath propagation can hinder direct recovery of the LoS parameters. In particular, the LoS and non-line-of-sight (NLoS) components associated with the same source are driven by the same transmitted signal and are therefore highly coherent, which can distort the LoS-induced signal subspace~\cite{wen2019survey}. Existing approaches typically first estimate the parameters of multiple propagation paths, followed by path association to determine which estimated paths correspond to the same source~\cite{xie2017source,liu2023iterative}. However, such path grouping does not directly identify the LoS paths of all sources, and errors in path estimation and association may further affect subsequent LoS-path identification.
	
	Third, existing near-field localization methods~\cite{yang2025beamforming,cheng2022efficient,rinchi2022cs,xu2025userlocalizationchannelestimation,challa1995high,yuen2002performance1,zuo2018new,huang2002near,zhi2007near,zhang2018rdmusic,gui_2020_cumulant,qu2024two,su2021cnn_near,Gast2026NFsubspace,gast2025dcd,zhu2025transformer} typically provide only point estimates, without explicit finite-sample coverage guarantees for source positions under multipath propagation, noise, finite observations, and model imperfections. Although reference~\cite{khurjekar2023doa} applied CP to DoA uncertainty quantification in far-field scenarios, simultaneous coverage guarantees for unordered multi-source position estimates in near-field localization remain largely unexplored. Consequently, it remains unclear how reliably the estimated positions represent the true source locations.
	
	Motivated by these challenges, this paper investigates reliable near-field multi-source enumeration and positioning in coherent multipath environments, with the main goal of achieving direct source positioning without estimating NLoS-path parameters or performing subsequent path association, while providing explicit reliability guarantees for both source enumeration and position estimation. The main contributions are as follows:
	
	
	\begin{itemize}
		\item \revision{We develop a three-stage framework for reliable near-field multi-source localization under coherent multipath propagation. The framework combines CRC-calibrated source enumeration, MUSIC-Net-based direct LoS positioning, and split conformal prediction (SCP)-based positioning calibration to meet the reliability requirements for source enumeration and positioning.}
		
		\item We propose a CRC-calibrated source enumeration method. We formulate eigenvalue-based source enumeration using a unified penalized-likelihood criterion and establish the monotonicity of the MDR loss with respect to its penalty coefficient. Based on this property, we calibrate the penalty coefficient via CRC to control the expected MDR at a prescribed level.
		
		\item We propose \emph{MUSIC-Net}, a two-stage MUSIC informed positioning framework for direct near-field LoS positioning in coherent multipath environments. By employing subspace-distance-based training objectives to isolate the LoS-related signal subspace and learn the surrogate distance required by two-stage MUSIC, MUSIC-Net directly estimates sources' LoS positions without explicit NLoS-path parameter estimation or subsequent path association.
		
		\item We develop an SCP-based positioning calibration method that separately calibrates the DoA and distance estimation errors to construct position predicted regions with simultaneous coverage guarantees for all sources. 
		
		\item Numerical results show that  the CRC-calibrated enumeration method achieves the prescribed MDR control, MUSIC-Net improves positioning accuracy in coherent multipath environments, and the SCP-based calibration method attains the target simultaneous coverage while constructing efficient position predicted regions.
	\end{itemize}
	
	\revision{The remainder of this paper is organized as follows. Section~\ref{sec:System Model and Problem Description} presents the system model, reliability requirements, and framework overview. The source-enumeration, positioning, and calibration stages are developed in Sections~\ref{sec:reliable_enumeration}, \ref{sec:subs_NN}, and~\ref{sec:conformal_calibration}, respectively. Numerical results are presented in Section~\ref{sec:numerical_results}, followed by conclusions in Section~\ref{sec:conclusion}.}

	\emph{Notations:}
	Scalars, vectors, and matrices are denoted by italic letters, boldface lower-case letters, and boldface upper-case letters, respectively. The sets of real and complex numbers are denoted by $\mathbb R$ and $\mathbb C$, respectively. The superscripts $(\cdot)^T$ and $(\cdot)^H$ denote transpose and Hermitian transpose, respectively. The notation $\boldsymbol I$ denotes an identity matrix with proper dimension, and $\boldsymbol 0$ denotes an all-zero vector or matrix with proper dimension. For a positive integer $A$, $[A]$ denotes the set $\{1,\ldots,A\}$. For a real number $A$, $\lceil A\rceil$ denotes the smallest integer greater than or equal to $A$. The notation $\|\cdot\|_2$ denotes the Euclidean norm, and $|\mathcal A|$ denotes the cardinality of a finite set $\mathcal A$. The operators $\operatorname{tr}(\cdot)$, $\operatorname{det}(\cdot)$, and $\operatorname{rank}(\cdot)$ denote the trace, determinant, and rank of a matrix, respectively. The operators $\mathbb E[\cdot]$ and $\mathbb P(\cdot)$ denote expectation and probability, respectively. The notation $\mathcal{CN}(\boldsymbol\mu,\boldsymbol\Sigma)$ denotes the circularly symmetric complex Gaussian distribution with mean $\boldsymbol\mu$ and covariance $\boldsymbol\Sigma$. The notation $\sigma_k(\boldsymbol A)$ denotes the $k$-th largest singular value of matrix $\boldsymbol A$. Finally, $j=\sqrt{-1}$ denotes the imaginary unit.

	\vspace{2ex}
	\section{{System Model and Problem Formulation}}\label{sec:System Model and Problem Description}
	\subsection{System Model}
	\label{subsec:system_model}
\begin{figure}[t]
		\centering
		\includegraphics[width=3in]{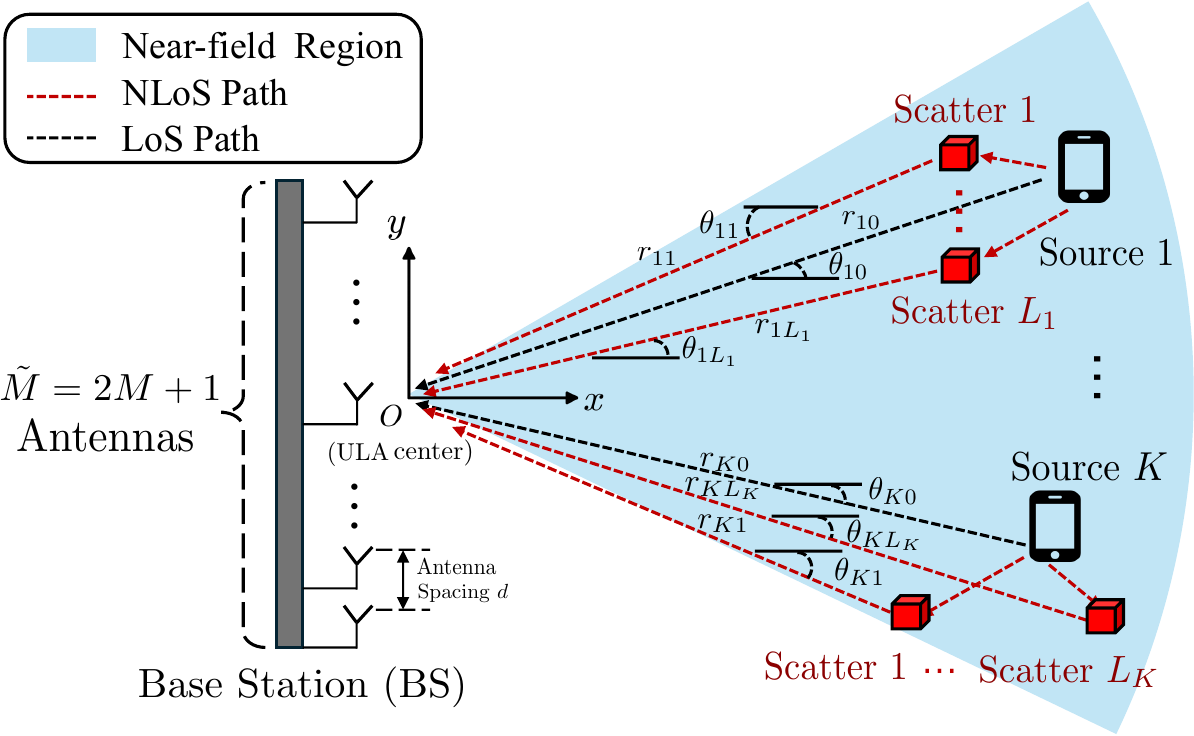}
		\caption{An illustration of the system for near-field multi-source localization with multiple coherent NLoS paths.}
		\label{fig:system_model}
	\end{figure}

	As shown in Fig.~\ref{fig:system_model}, we consider a multi-source narrowband near-field system for uplink localization, in which a base station (BS) employs a uniform linear array (ULA) with $\tilde{M}=2M+1$ antennas. The system contains an unknown number $K$ of single-antenna sources with $K\in [\tilde M-1]$. All sources are assumed to lie in the near-field region, i.e., the distance $r$ from each source to the center of the ULA satisfying $r<2D^2/\lambda$, where $D=(\tilde{M}-1)d$ is the array aperture; $d$ is the antenna spacing; and $\lambda$ is the carrier wavelength. 
	
	\label{sec:system_model_problem_formulation}
	
	Under the narrowband near-field assumption, the steering vector $\boldsymbol a(\theta,r)\in\mathbb{C}^{\tilde{M}\times 1}$ for a target source at position \(\boldsymbol{p}=r[\cos\theta,\sin\theta]^T\) is given by \cite{liu2023near}
	\begin{equation}
		\boldsymbol a(\theta,r)=
		\left[
		e^{-j\frac{2\pi}{\lambda}\|\boldsymbol p-\boldsymbol s_{-M}\|_2},
		\ldots,
		e^{-j\frac{2\pi}{\lambda}\|\boldsymbol p-\boldsymbol s_{M}\|_2}
		\right]^T,
		\label{eq:steering_vector}
	\end{equation}
	where $\boldsymbol s_n=[0,nd]^T$ denotes the position of the 
	antenna element \(n\in\{-M,-M+1, \ldots, M-1,M\}\); $\theta\in[-\pi/2,\pi/2]$ and
	$r\in[r_{\min},r_{\max}]$ denote the DoA and distance of the target source,
	respectively.
	We assume that, for the signal transmitted by the $k$-th source, there exists one LoS propagation path and $L_k\ge 0$ NLoS ones generated by scatterers between the $k$-th source and the BS. Let $s_k(t)$ denote the signal transmitted by source $k\in[K]$ at time slot $t\in[T]$, where $T$ is the total number of time slots with $T>K$. For each source $k \in [K]$,  $s_k(t)\sim\mathcal{CN}(0,P_k)$ are independent across sources and time slots,   where $P_k$ denotes  the transmit power of source $k$. 
	
	The received signal $\mv y(t) \in \mathbb{C}^{\tilde M \times 1}$ can thus be expressed as
	\begin{multline}
		\boldsymbol y(t)
		= \underbrace{\sum_{k=1}^{K}
			\boldsymbol a(\theta_{k0},r_{k0})s_k(t)
		}_{\boldsymbol y_{\mathrm{LoS}}(t)}
		\\
		+ \underbrace{\sum_{k=1}^{K}\sum_{l=1}^{L_k}
			\boldsymbol a(\theta_{kl},r_{kl})c_{kl}s_k(t)
		}_{\boldsymbol y_{\mathrm{NLoS}}(t)}
		+ \boldsymbol n(t),
		\label{eq:received_signal}
	\end{multline}
	where 
	$\theta_{kl}$ and $r_{kl}$ denote the DoA and propagation distance of the $l$-th path associated with the $k$-th source , respectively, with $l=0$ being the index of the LoS path; $c_{kl}$ is the complex reflection coefficient of the $l$-th NLoS path of the $k$-th source; and $\boldsymbol n(t)\sim \mathcal{CN}(\boldsymbol 0,\sigma^2\boldsymbol I)$ is the additive white Gaussian noise (AWGN) received at the BS. Accordingly, the covariance matrix of $\boldsymbol y(t)$ is given by
	\begin{align}
		\begin{aligned}
			\boldsymbol R & = \mathbb{E} [\boldsymbol y(t)\boldsymbol y^H(t)]
			\\	&= \underbrace{ \mathbb{E}[\boldsymbol y_{\mathrm{LoS}}(t)\boldsymbol y_{\mathrm{LoS}}^H(t)]}_{\boldsymbol R_{\mathrm{LoS}}} + \underbrace{\mathbb{E}[\boldsymbol y_{\mathrm{NLoS}}(t)\boldsymbol y_{\mathrm{NLoS}}^H(t)]}_{\boldsymbol R_{\mathrm{NLoS}}} \\& + \underbrace{\mathbb{E}[\boldsymbol y_{\mathrm{LoS}}(t)\boldsymbol y_{\mathrm{NLoS}}^H(t)] + \mathbb{E}[\boldsymbol y_{\mathrm{NLoS}}(t)\boldsymbol y_{\mathrm{LoS}}^H(t)]}_{\boldsymbol R_{\mathrm{cross}} } +\sigma^2\boldsymbol I.
		\end{aligned}
		\label{eq:all_covariance}
	\end{align}
	In practice,  the  covariance matrix $\boldsymbol{ R}$ can be approximated by the  sample covariance matrix (SCM)   of the observations
	\begin{equation}
		\boldsymbol{\widehat R} = \frac{1}{T} {\boldsymbol Y} {\boldsymbol Y}^H, \label{eq:SCM}
	\end{equation}
	where $\boldsymbol Y=[\boldsymbol y(1),\ldots,\boldsymbol y(T)] \in \mathbb{C}^{\tilde M \times T}$.

	\subsection{Problem Formulation}
	\label{subsec:joint_estimation}
	Given the observations in \eqref{eq:received_signal}, our goal is to provide reliable estimation of the
	unknown number $K$ of  sources and their LoS location-related parameters $\boldsymbol{\Theta}_{K} = \{\theta_{k0},r_{k0}\}_{k=1}^{K}$. 
	
	A natural starting point is to maximize the likelihood of the received observations. 
	Under the complex Gaussian model \(\boldsymbol{y}(t)\sim\mathcal{CN}(\boldsymbol{0},\boldsymbol{R}(\boldsymbol{\Xi}_K,K))\), maximizing the log-likelihood $\ln p(\boldsymbol Y;\boldsymbol \Xi_K,K)$ over the unknown parameters is equivalent to solving
	\begin{equation}
		\begin{aligned}
			\mathop{\mathtt{Minimize}}_{\boldsymbol{\Xi}_K,K}\quad
			\ln\det\!\bigl(\boldsymbol R(\boldsymbol\Xi_K,K)\bigr)
			+\operatorname{tr}\!\left(
			\boldsymbol R(\boldsymbol\Xi_K,K)^{-1}\widehat{\boldsymbol R}
			\right),
		\end{aligned}
		\label{eq:joint_estimation}
	\end{equation}
	where $\boldsymbol{\Xi}_K\triangleq \{\boldsymbol{\Theta}_{K},\boldsymbol{\Psi}_{K}, \sigma^2\}$ collects all unknown parameters:  $\boldsymbol{\Psi}_{K} = \{\boldsymbol{\Phi}_k\}_{k=1}^K$ with  $\boldsymbol{\Phi}_k= \{\theta_{kl},r_{kl},c_{kl}\}_{l=1}^{L_k}$ denotes the parameters associated with the NLoS paths, which are treated as nuisance parameters;  $\boldsymbol R(\boldsymbol{\Xi}_K,K)$ denotes the covariance matrix in \eqref{eq:all_covariance} parameterized by $\boldsymbol{\Xi}_K$ and $K$.

	
	\begin{figure*}[t]
		\centering
		\includegraphics[width=7.2in]{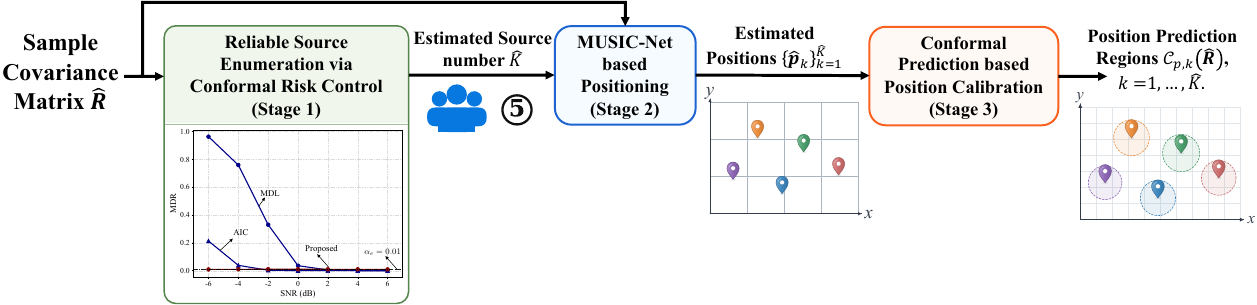}
		\caption{The overall protocol of the proposed reliable multi-source enumeration and localization framework. The plot in Stage~1 shows the
			MDR versus SNR for AIC, MDL, and the proposed CRC-calibrated scheme.
			The dashed line indicates the target MDR level $\alpha_e=0.01$.} 
		\label{fig:proposed_work_flow}
	\end{figure*}

	\begin{figure*}[t]
		\centering
		\includegraphics[width=6in]{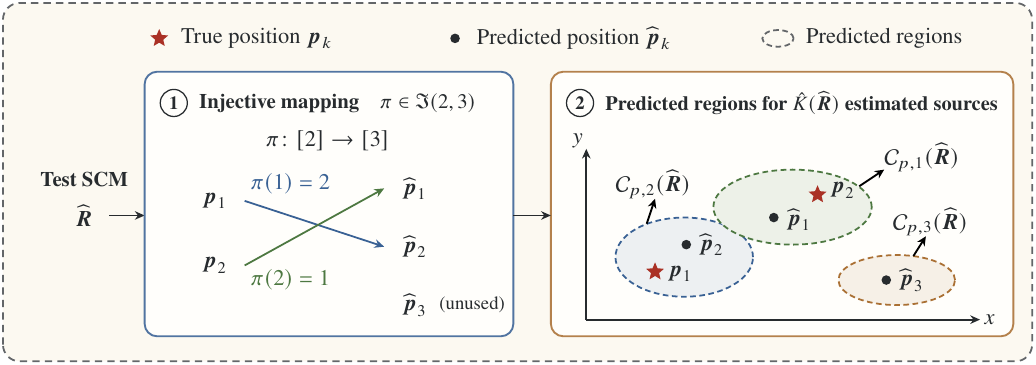}
		\caption{Illustration of an injective mapping and the associated predicted regions with $K=2$ and $\widehat{K}[\widehat{\boldsymbol{R}}]=3$.}
		\label{fig:position_coverage_matching}
	\end{figure*}
	
	First, to address uncertainty attributed by source enumeration in safety-critical applications, such as autonomous driving and unauthorized unmanned aerial vehicle (UAV)  detection, source underestimation is generally more consequential than overestimation because missed sources are excluded from subsequent provisioning and monitoring \cite{khawaja2025survey,wang2021counter}, thus causing potentially serious safety concerns. We  therefore prioritize achieving low MDR over low false-alarm rate (FAR), imposing the following empirical risk-control requirement
	\begin{equation}
		\mathbb E\!\left[\ell_{\mathrm{MDR}}(\widehat K,K)\right]\le \alpha_e,
		\label{eq:reliable_source_enum_goal}
	\end{equation}
	where $\alpha_e\in(0,1)$ is a prescribed MDR level; and $\ell_{\mathrm{MDR}}(\widehat K,K)$ denotes the MDR loss, given by
	\begin{equation}
		\ell_{\mathrm{MDR}}(\widehat K,K)
		=
		\frac{\max(K-\widehat K,0)}{K},
		\label{eq:source_under_loss}
	\end{equation}
	where $\widehat K$ represents the  estimated number of sources; and the expectation in \eqref{eq:reliable_source_enum_goal} is taken with respect  to all data samples $\hat{\mv R}$.

	\begin{remark}
		However, the standard formulations of AIC and MDL, two widely used source enumeration criteria~\cite{stoica2004model}, which adopt the ITC-based approach that balances the maximized log-likelihood against a penalty for model complexity, do not nevertheless provide in general finite-sample MDR guarantee required by \eqref{eq:reliable_source_enum_goal}. For example, given a target MDR level $\alpha_e=0.01$, Fig.~\ref{fig:proposed_work_flow} shows how MDRs achieved by AIC and MDL vary with SNR, both possibly exceeding $\alpha_e$ in low SNR regime. 
	\end{remark}

	Next, to accommodate uncertainty arising from coherent multipath propagation, AWGN, and limited observations, we aim for providing predicted region estimates with a prescribed confidence level, rather than point estimates alone.
	
	Specifically, define the true and estimated position  by
	$\boldsymbol p_k=r_{k0}[\cos(\theta_{k0}),\sin(\theta_{k0})]^T$ and $\widehat{\boldsymbol p}_k=\widehat{r}_{k0}[\cos(\widehat{\theta}_{k0}),\sin(\widehat{\theta}_{k0})]^T$, respectively.
	Let $\widehat K (\widehat{\boldsymbol{R}})$ denote the estimated number of sources returned by any joint source enumeration and position estimator for input $\widehat{\boldsymbol{R}}$. Let $\mathfrak I(K,\widehat{K}(\widehat{\boldsymbol{R}})$ denote a set of injective mappings \(\{\pi: [K]\to[\widehat K(\widehat{\mv R})] \left.|\right. \forall k\in[K], \exists \pi(i)\in[\widehat K(\widehat{\mv R})]\}\).
	Our goal is to construct  an injective mapping from $\mathfrak I(K,\widehat{K}(\widehat{\boldsymbol{R}}))$ and  $\widehat{K}(\widehat{\boldsymbol{R}})$ predicted regions
	$\mathcal C_{p,j}(\widehat{\boldsymbol{R}})$, $j \in [\widehat{K}(\widehat{\boldsymbol{R}})]$, such that all regions satisfy the following  coverage requirement:
	\begin{align}
		\mathbb P\Bigl(
		\boldsymbol p_1\in\mathcal C_{p,\pi(1)}(\widehat{\boldsymbol{R}}),\ldots, \boldsymbol p_K\in\mathcal C_{p,\pi(K)}(\widehat{\boldsymbol{R}})	\Bigr)
		\geq 1-\alpha_p,
		\label{eq:joint_position_coverage_requirement}
	\end{align}
	where $\pi(k)$ denotes the index in the prediction region assigned to the true source $k$ under the injective matching $\pi$; $\alpha_p\in(0,1)$ is the target miscoverage level for localization; and the probability is taken with respect to the predicted regions built upon data samples, which will be elaborated in later sections.  As illustrated in Fig.~\ref{fig:position_coverage_matching}, proposed construction provides each true source $k\in [K]$ with a region estimate $\mathcal{C}_{p,\pi(k)}(\widehat{\mv R})$ indexed by a distinct \(\pi(k)\in[\hat K(\widehat{\mv R})]\).

	To sum up, the reliability-constrained joint multi-source enumeration and localization problem is formulated as follows:
	\begin{equation}
		\begin{aligned}
			&\underset{\widehat{\boldsymbol\Xi}_K,\{\mathcal{C}_{p,\pi(k)}\}_{k=1}^K,\widehat{K}}{\mathtt{Minimize}}
			\ln\det\!\bigl(\boldsymbol R(\widehat{\boldsymbol\Xi}_K,\widehat{K})\bigr)
			+\operatorname{tr}\!\left(
			\boldsymbol R(\widehat{\boldsymbol\Xi}_K,\widehat{K})^{-1}\widehat{\boldsymbol R}
			\right)
			\\
			&~~\mathtt{subject\ to}\quad ~~
			\eqref{eq:reliable_source_enum_goal}
			\ \text{and}\ 
			\eqref{eq:joint_position_coverage_requirement}.
		\end{aligned}
		\label{eq:reliability_constrained_joint_problem}
	\end{equation}
	The objective function is the log-likelihood of the observed data  (c.f.~\eqref{eq:joint_estimation}), subject to the empirical risk control over the MDR and the marginal confidence level for a true target being covered by the predicted region.

	As problem \eqref{eq:reliability_constrained_joint_problem} is highly intractable due to the unknown integer number $K$ of sources and its coupling with the LoS and nuisance parameters, we propose in this paper a novel data-driven solution to provide joint multi-source enumeration and localization that meets the reliability constraints in \eqref{eq:reliable_source_enum_goal} and 	\eqref{eq:joint_position_coverage_requirement}. The proposed framework is summarized in Fig.~\ref{fig:proposed_work_flow}, where related neural network training and statistical calibration are all performed offline. At inference, given a sample SCM $\boldsymbol{\widehat R}$, the framework proceeds as follows. \textbf{Stage~1} (c.f.~Section \ref{sec:reliable_enumeration}): CRC based source enumeration returns the estimated source number $\widehat K[\widehat{\boldsymbol{R}}]$. \textbf{Stage~2} (c.f.~Section~\ref{sec:subs_NN}): MUSIC-Net based positioning   uses $(\widehat{\boldsymbol{R}},\widehat K[\widehat{\boldsymbol{R}}])$ as input to obtain a number $\widehat K[\widehat{\boldsymbol{R}}]$ of point position estimates $\{\widehat{\boldsymbol p}_k\}_{k=1}^{K[\widehat{\boldsymbol{R}}]}$. \textbf{Stage~3} (c.f.~Section~\ref{sec:conformal_calibration}): Post-hoc calibration leveraging SCP builds upon these estimates to construct the predicted regions $\{\mathcal C_{p,\pi(k)}(\widehat{\boldsymbol{R}})\}_{k=1}^{K}$.

	\section{\revision{Stage 1: CRC Based Source Enumeration}}
	\label{sec:reliable_enumeration}
	{In this section, we develop a CRC based source-enumeration method to meet the expected MDR requirement in \eqref{eq:reliable_source_enum_goal}. Given a test SCM $\widehat{\boldsymbol R}$, the calibrated criterion returns $\widehat K$, which determines the noise-subspace dimension $\tilde M-\widehat K$ and the number of position estimates in Stage~2.}

	\subsection{Preliminary: Conformal Risk Control~\cite{angelopoulos_conformal_2024}}
	\label{subsec:crc_prelim}
	CRC  is a distribution-free calibration framework that provides finite-sample guarantees on the expected value of a task-dependent loss, such as the false negative rate (FNR) or other bounded task-specific losses. Let $\mathcal D_{\mathrm{cal}}=\{(X^{(i)},Y^{(i)})\}_{i=1}^{n}$ be a calibration dataset of $n$  independent and identically distributed (i.i.d.) samples from a distribution over $\mathcal X\times\mathcal Y$, where $X^{(i)}$ is the feature and $Y^{(i)}$ is the corresponding label.
	Let $\{\mathcal C_\lambda(\cdot):\lambda\in\Lambda\}$ denote a family of set-valued predictors indexed by a scalar calibration parameter $\lambda$, where $\Lambda\subseteq\mathbb R$ is the search space. For a bounded task-specific loss function $\ell(\mathcal C_\lambda(x),y)\le B$, with \(B<\infty\), CRC selects a data-dependent parameter $\widehat\lambda$ such that
	\begin{equation}
		\mathbb E\!\left[
		\ell\!\left(\mathcal C_{\widehat\lambda}(X^{(n+1)}),Y^{(n+1)}\right)
		\right]
		\le \beta,
		\label{eq:crc_goal}
	\end{equation}
	where $\beta<B$ is a prescribed risk level and the expectation is taken over both the calibration data and the test sample.
	
	For each $\lambda\in\Lambda$, the empirical risk on the calibration set is given by
	\begin{equation}
		\widehat R(\lambda)
		=
		\frac{1}{n}
		\sum_{i=1}^{n}
		\ell\!\left(
		\mathcal C_\lambda(X^{(i)}),Y^{(i)}
		\right).
		\label{eq:crc_empirical_risk}
	\end{equation}
	CRC then selects the calibration parameter as
	\begin{equation}
		\widehat\lambda
		=
		\inf
		\left\{
		\lambda\in\Lambda:
		\widehat R(\lambda)
		\le
		\beta-\frac{B-\beta}{n}
		\right\}.
		\label{eq:crc_lambda_hat}
	\end{equation}
	If the calibration samples in $\mathcal{D}_{\text{cal}}$ and test data sample $(X^{(n+1)}, Y^{(n+1)})$ are exchangeable, and \(\ell(\mathcal{C}_\lambda(X^{(i)}),Y^{(i)})\) is bounded and pointwise non-increasing in \(\lambda\), then  \eqref{eq:crc_lambda_hat} guarantees \eqref{eq:crc_goal}~\cite[Theorem~1]{angelopoulos_conformal_2024}.

	\subsection{Penalized-Likelihood Source Enumeration}
	From \eqref{eq:received_signal}, all propagation paths associated with the same
	source are driven by the same transmitted signal $s_k(t)$. Therefore,
	\eqref{eq:received_signal} can be equivalently rewritten as
	\begin{equation}
		\boldsymbol y(t)
		=
		\sum_{k=1}^{K}
		\boldsymbol b_k s_k(t)
		+
		\boldsymbol n(t),
	\end{equation}
	where
	\begin{equation}
		\boldsymbol b_k
		=
		\boldsymbol a(\theta_{k0},r_{k0})
		+
		\sum_{l=1}^{L_k}
		c_{kl}\boldsymbol a(\theta_{kl},r_{kl})
	\end{equation}
	denotes the equivalent steering vector of source $k$.
	Since the source signals $\{s_k(t)\}_{k=1}^{K}$ are mutually independent,
	the covariance matrix in \eqref{eq:all_covariance} can be rewritten as
	\begin{equation}
		\boldsymbol R
		=
		\sum_{k=1}^{K}
		P_k
		\boldsymbol b_k\boldsymbol b_k^H
		+
		\sigma^2\boldsymbol I.
	\end{equation}
	Thus, the covariance matrix $\boldsymbol R$ comprises $K$ source-dependent rank-one components and spatially white noise. Accordingly, source enumeration can be performed by applying eigenvalue-based information criteria, such as MDL and AIC, to the SCM $\widehat{\boldsymbol R}$, without resorting to additional coherent-path separation techniques \cite{xie2017source}.
	We express these eigenvalue-based information criteria in
	a unified penalized-likelihood representation 
	\begin{align}
		\widehat K_\eta(\widehat{\boldsymbol R})
		=
		\arg \min _{q \in\{0, \ldots, \tilde{M}-1\}}
		\left[
		-2 \ln p\left(\boldsymbol{Y} ; \widehat{\boldsymbol{\Xi}}_q,q\right)
		+ G(q)
		\eta
		\right],
		\label{eq:source_number_selection}
	\end{align}
	where $G(q)=\bigl(q(2 \tilde{M}-q)+1\bigr) $; $q$ denotes a candidate source number, with the convention $[0]=\varnothing$; $\widehat{\boldsymbol{\Xi}}_q$ denotes the maximum likelihood estimates of the unknown parameters under the hypothesis that the source number is $q$; $\eta \ge 0$ is the penalty coefficient, with $\eta=2$ yielding AIC and $\eta=\ln T$ yielding MDL; and $\ln p(\boldsymbol Y;\widehat{\boldsymbol{\Xi}}_q,q)$ is the  log-likelihood of the observation  $\boldsymbol Y$ under the hypothesis that the source number is $q$, whose closed form is given by \cite{akaike1998information}
	\begin{equation}
		\ln p\left(\boldsymbol{Y};\widehat{\boldsymbol{\Xi}}_q,q\right)
		=
		- T(\tilde M-q)\log\left(
		\frac{
			\frac{1}{\tilde M-q}\sum_{m=q+1}^{\tilde M}\widehat\lambda_m
		}{
			\left(\prod_{m=q+1}^{\tilde M}\widehat\lambda_m\right)^{\frac{1}{\tilde M-q}}
		}
		\right),
		\label{eq:mdl_likelihood}
	\end{equation}
	where $\widehat\lambda_m$ denotes the $m$-th largest eigenvalue of the SCM $\widehat{\boldsymbol R}$.

	\revision{The penalty coefficient $\eta$ balances likelihood fit against model complexity~\cite{stoica2004model}. Larger values favor smaller estimates of the source number, while smaller values reduce the penalty on additional sources. We next calibrate $\eta$ using CRC to meet the expected MDR requirement in \eqref{eq:reliable_source_enum_goal}.}

	\subsection{CRC based Source Enumeration}
	\label{subsec:CRC_source_enum}
	
	To meet the reliable source enumeration requirement in \eqref{eq:reliable_source_enum_goal}, we calibrate the penalty coefficient $\eta$ in \eqref{eq:source_number_selection} via CRC. Let
	$\mathcal D_{\mathrm{cal}}^{\mathrm{se}}
	=\{(\widehat{\boldsymbol R}^{(i)},K^{(i)})\}_{i=1}^{n}$
	denote the calibration dataset for source enumeration, where $K^{(i)}$ is the true number of sources of the $i$-th calibration sample. We assume that the calibration samples in $\mathcal D_{\mathrm{cal}}^{\mathrm{se}}$ and a future test sample $(\widehat{\boldsymbol R},K)$ are exchangeable.
	
	For each calibration sample $i\in[n]$ and a given penalty coefficient $\eta$, let $\widehat K^{(i)}(\eta)=\widehat K_\eta(\widehat{\boldsymbol R}^{(i)})$ denote the corresponding source-number estimate from \eqref{eq:source_number_selection}. Using the MDR in \eqref{eq:source_under_loss} as the sample-wise loss, we compute the empirical calibration risk as
	\begin{equation}
		\widehat R(\eta)
		=
		\frac{1}{n}
		\sum_{i=1}^{n}
		\ell_{\mathrm{MDR}}
		\left(
		\widehat K^{(i)}(\eta),K^{(i)}
		\right).
		\label{eq:source_enum_empirical_risk}
	\end{equation}
	
	Applying CRC requires a monotonic relationship between the calibration parameter and the loss. The following lemma establishes this relationship for $\eta$.
	
	\begin{lemma}[Monotonicity of MDR loss in \eqref{eq:source_under_loss}]
		\label{lem:source_loss_monotone}
		For any fixed SCM $\widehat{\boldsymbol R}$, the source-number estimate 
		$\widehat K_\eta(\widehat{\boldsymbol R})$ obtained from \eqref{eq:source_number_selection} can be chosen to be monotone non-increasing in $\eta$. Consequently, for any true source number $K\ge 1$, the MDR loss in
		\eqref{eq:source_under_loss} is monotone non-decreasing in $\eta$.
	\end{lemma}
	
	\begin{IEEEproof}
		The proof is provided in Appendix~\ref{app:proof_source_loss_monotone}.
	\end{IEEEproof}
	By Lemma~\ref{lem:source_loss_monotone}, the MDR loss in \eqref{eq:source_under_loss} is monotone non-decreasing in $\eta$. We therefore set $\lambda=-\eta$ for which the loss is non-increasing. Following the CRC construction~\cite{angelopoulos_conformal_2024} with \eqref{eq:crc_lambda_hat} with $B=1$ and target risk $\beta=\alpha_e$, the calibrated penalty coefficient is 
	\begin{equation}
		\widehat \eta
		=
		\sup
		\left\{
		\eta\ge 0:
		\widehat R(\eta)
		\le
		\alpha_e-\frac{1-\alpha_e}{n}
		\right\}.
		\label{eq:crc_eta_hat}
	\end{equation}
	If the feasible set in \eqref{eq:crc_eta_hat} is empty, we set $\widehat\eta=0$, since it corresponds to the weakest penalty and thus favors overestimation rather than miss detection \cite{angelopoulos_conformal_2024}.
	
	For a test SCM $\widehat{\boldsymbol R}$, we estimate the number of sources by solving \eqref{eq:source_number_selection} with $\eta=\widehat\eta$, yielding $\widehat K=\widehat K_{\widehat\eta}(\widehat{\boldsymbol R})$. Under exchangeability of the calibration and test samples and by the monotonicity of the MDR loss with respect to $\lambda=-\eta$, the CRC guarantee in Section~\ref{subsec:crc_prelim} ensures that this estimate satisfies \eqref{eq:reliable_source_enum_goal}.

	\begin{figure*}[h]
		\centering
		\includegraphics[width=6in]{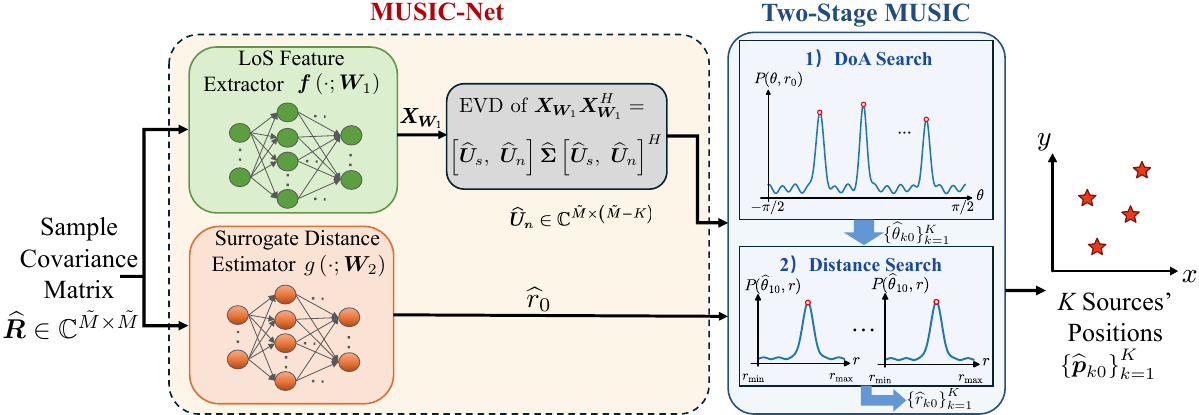}
		\caption{The proposed two-stage MUSIC informed multi-source positioning based on MUSIC-Net.}
		\label{fig:proposed_architecture}
	\end{figure*}

	\section{\revision{Stage 2: MUSIC-Net-Based Direct LoS Positioning}}
	\label{sec:subs_NN}
	\revision{In this section, we propose MUSIC-Net, a model-based DL framework for direct LoS positioning using the SCM $\widehat{\boldsymbol R}$ and the estimated source number $\widehat K$ from Stage~1. MUSIC-Net learns the LoS-related subspace and surrogate distance required by two-stage MUSIC~\cite{qu2024two}.}
	\subsection{{Two-Stage MUSIC}}
	In the standard two-stage MUSIC algorithm \cite{qu2024two},  since the
	LoS paths of the sources dominate the  covariance matrix $\boldsymbol{R}$, the NLoS and cross-correlation terms $\boldsymbol R_{\mathrm{NLoS}}+\boldsymbol R_{\mathrm{cross}}$
	in \eqref{eq:all_covariance} are assumed to be negligible. Accordingly, $\boldsymbol{R}$ in \eqref{eq:all_covariance} can be approximated as  \(\boldsymbol{R}=\boldsymbol A_{\mathrm{LoS}} \operatorname{diag}(P_1,\ldots,P_K)\boldsymbol A_{\mathrm{LoS}}^H  +\sigma^2\boldsymbol{I}\), where $\boldsymbol A_{\mathrm{LoS}} = [\boldsymbol{a}(\theta_{10},r_{10}),\ldots,\boldsymbol{a}(\theta_{K0},r_{K0})]$. An eigenvalue decomposition (EVD) of $\mv R$ then decomposes its eigenspace into a signal subspace $\mv U_s$ and a noise subspace $\mv U_n$ as assumed in the conventional far-field MUSIC algorithm.
	MUSIC algorithm for the near-field scenario performs a two-dimension search for the $K$ largest peaks based on the spectrum function given by
	\begin{align}
		P(\theta, r)= \frac{1}{\|\boldsymbol{a}^H \left( \theta,r\right) \boldsymbol{U}_n\|_2}. \label{eq:hehe}
	\end{align}
	Instead of an exhaustive 2D search over $[-\tfrac{\pi}{2}, \tfrac{\pi}{2}]\times[r_{\min}, r_{\max}]$, two-stage MUSIC decouples the above search into two one-dimension spectral searches, and performs them in the following two stages. 1) \textbf{DoA search}: Based on a surrogate distance $r_0$, which is obtained by either  maximizing the minimum correlation (MMC) rule that maximizes the worst-case steering vector correlation or exceeding the minimum correlation threshold (EMCT) rule that enforces this correlation to exceed a prescribed threshold \cite{qu2024two}, estimate the DoAs of the sources by searching for the $K$ peaks with $r$ replaced by $r_0$ in \eqref{eq:hehe}; and 2) \textbf{Distance search}: Based on the estimated DoAs $\widehat{\theta}_{k0}$, $k\in[K]$, estimate the distance corresponding to each source $k\in[K]$ by searching for one peak with $\theta$ replaced by $\widehat{\theta}_{k0}$ in \eqref{eq:hehe}.

	Under coherent multipath propagation, however, the NLoS and cross-correlation terms in \eqref{eq:all_covariance} distort the LoS-induced subspace, and the noise subspace required by conventional two-stage MUSIC can no longer be reliably recovered directly from the SCM. This motivates the data-driven design introduced next.


	\subsection{MUSIC-Net Training}
	MUSIC-Net augments two-stage MUSIC with two trainable modules: a LoS feature extractor for recovering the MUSIC noise subspace \(\mv U_n\) and a surrogate-distance \(r_0\) estimator for the angular search. As illustrated in Fig. \ref{fig:proposed_architecture}, MUSIC-Net  consists of 1) a LoS feature extractor module and 2) a surrogate distance estimation module, which produce the estimated LoS-plus-noise subspace  $\widehat{\mv U}_{\!n}$ and a \emph{surrogate distance} estimation $\widehat{r}_0$, respectively, from the input SCM $\widehat{\boldsymbol R}$.

	Specifically,  the LoS feature extractor, denoted by $\boldsymbol f(\cdot;\boldsymbol W_{\!1})$ with trainable parameters $\boldsymbol W_{\!1}$, maps SCM \( \widehat{\boldsymbol R}\) to a complex-valued feature matrix \(\boldsymbol X_{\!\boldsymbol W_{\!1}}\in\mathbb C^{\tilde M\times\tilde M}\), i.e., $\boldsymbol f(\widehat{\boldsymbol R};\boldsymbol W_{\!1})=\boldsymbol X_{\!\boldsymbol W_{\!1}}$. It is trained so that the dominant eigenspace of $\boldsymbol X_{\!\boldsymbol W_{\!1}}\boldsymbol X_{\!\boldsymbol W_{\!1}}^H$ aligns with the signal subspace dominated by the LoS paths, thereby providing a learned surrogate for the noise-subspace input required by conventional MUSIC based algorithms.
	Then, we perform the EVD on $\boldsymbol X_{\!\boldsymbol W_{\!1}}\boldsymbol X_{\!\boldsymbol W_{\!1}}^H$:
	\begin{equation}
		\boldsymbol X_{\!\boldsymbol W_{\!1}} \boldsymbol X_{\!\boldsymbol W_{\!1}}^H
		=
		\left[
		\widehat{\boldsymbol U}_{\! s},~
		\widehat{\boldsymbol U}_{\!n}
		\right]
		\widehat{\boldsymbol \Sigma}
		\left[
		\widehat{\boldsymbol U}_{\!s},~
		\widehat{\boldsymbol U}_{\!n}
		\right]^H,
		\label{eq:evd_of_surrogate}
	\end{equation}
	where $\widehat{\boldsymbol U}_{\! s}\in\mathbb C^{\tilde M\times K}$ contains the eigenvectors associated with the $K$ largest eigenvalues spanning over the estimated LoS-induced signal subspace; $\widehat{\boldsymbol U}_{\!n}\in\mathbb C^{\tilde M\times(\tilde M-K)}$ contains the remaining eigenvectors; and $\widehat{\boldsymbol \Sigma}$ is a diagonal matrix containing all eigenvalues in descending order. 
	Next, we adopt the subspace distance \cite{chen_2025_srl} between the target LoS-induced signal subspace $\mv U_{\!s}$, which is obtained by EVD on $\boldsymbol A_{\mathrm{LoS}}\boldsymbol A_{\mathrm{LoS}}^H$, and the learned LoS-induced signal subspace $\widehat{\mv U}_{\!s}$ as the supervised loss. Specifically, let $\mathcal D_{\mathrm{tr}}=\{(\widehat{\boldsymbol{R}}^{(i)},
	\boldsymbol{\Theta}_{K^{(i)}})\}_{i=1}^{n_{\mathrm{tr}}}$,
	where
	$n_{\mathrm{tr}}$ is the number of training
	samples,  $\boldsymbol{\Theta}_{K^{(i)}}=(\theta_{k0}^{(i)}, r_{k0}^{(i)})_{k=1}^{K^{(i)}}$, and $(\theta_{k0}^{(i)}, r_{k0}^{(i)})$ denotes the ground-truth DoA and distance of source $k\in[K^{(i)}]$ in the $i$-th sample. For each sample, let
	$\boldsymbol U_s^{(i)}$ and $\widehat{\boldsymbol U}_s^{(i)}$
	denote the $K^{(i)}$-dimension ground-truth LoS-induced signal subspace
	and its estimate, respectively. Then, the empirical training loss is given by
	\begin{align}
		\begin{aligned}
			\mathcal L_1 \left(\boldsymbol W_{\!1} \right)
			=
			\frac{1}{n_{\mathrm{tr}}}
			\sum_{i =1 }^{n_{\mathrm{tr}}}
			\sum_{k=1}^{K^{(i)}}
			\arccos\!\left(
			\sigma_k\!\left(\left( \boldsymbol U_s^{(i)}\right)^H\widehat{\boldsymbol U}_{\! s}^{(i)}\right)
			\right).
		\end{aligned}
		\label{eq:denoising}
	\end{align}
	The subspace distance directly targets the geometric quantity required by MUSIC and avoids reconstructing the full covariance. By contrast, covariance-reconstruction losses such as SCM fitting \cite{chen_2025_srl} and the affine-invariant distance \cite{barthelme2021doa} can be sensitive to the received SNR, potentially degrading generalization across SNR conditions. 
	
	The surrogate distance estimator, denoted by $g(\cdot;\boldsymbol W_{\!2})$ with trainable parameters $\boldsymbol W_{\!2}$, maps the SCM $\widehat{\boldsymbol R}$ to a surrogate distance $\widehat{r}_0$, i.e., $ \widehat{r}_0=g(\widehat{\boldsymbol R};\boldsymbol W_{\!2}).$ 
	To ensure that the surrogate LoS-induced subspace $\tilde{\mv U}_{\!s}$ associated with a single distance $\widehat r_0$, which is obtained by EVD on $\boldsymbol A_{\widehat{r}_0}\boldsymbol A_{\widehat{r}_0}^H$ with $\boldsymbol A_{\widehat{r}_0} =
	\big[ \boldsymbol a(\theta_{10},\widehat{r}_0),\ldots, \boldsymbol a(\theta_{K0},\widehat{r}_0) \big]$, aligns well with the estimated subspace $\widehat{\boldsymbol U}_{\!s}$ associated with the $K$ sources'  distances, the empirical training loss is therefore chosen as the subspace distance between $\widetilde{\boldsymbol U}_{\!s}$ and $\widehat{\boldsymbol U}_{\!s}$.  Specifically, given the $i$-th training sample $(\widehat{\boldsymbol{R}}^{(i)},
	\{(\theta_{k0}^{(i)},r_{k0}^{(i)})\}_{k=1}^{K^{(i)}})$, let
	$\widehat r_0^{(i)}
	=g(\widehat{\boldsymbol R}^{(i)};\boldsymbol W_{\!2})$ and
	$\widetilde{\boldsymbol U}_{\!s}^{(i)}$ denote the
	$K^{(i)}$-dimension dominant eigenspace obtained by applying EVD to
	$\boldsymbol A_{\widehat{r}_0}^{(i)}
	(\boldsymbol A_{\widehat{r}_0}^{(i)})^H$. Then, the empirical training loss is given by
	\begin{align}
		\begin{aligned}
			\mathcal L_2(\boldsymbol W_{\!1},\boldsymbol W_{\!2})
			=
			\frac{1}{n_{\mathrm{tr}}}
			\sum_{i=1}^{n_{\mathrm{tr}}}
			\sum_{k=1}^{K^{(i)}}
			\arccos\!\left(
			\sigma_k\!\left(
			\left(\widetilde{\boldsymbol U}_{\!s}^{(i)}\right)^H
			\widehat{\boldsymbol U}_{\!s}^{(i)}
			\right)
			\right).
		\end{aligned}
		\label{eq:virtual_distance_learning}
	\end{align}


	The two modules are trained jointly according to
	\begin{equation}
		\mathop{\mathtt{Minimize}}\limits_{\boldsymbol W_{\!1},\boldsymbol W_{\!2}}\;
		\mathcal L_1(\boldsymbol W_{\!1})+\mu\,\mathcal L_2(\boldsymbol W_{\!1},\boldsymbol W_{\!2}),
		\label{eq:joint_train}
	\end{equation}
	where $\mu>0$ balances the two losses.

	\begin{algorithm}[t]
		\caption{\revision{MUSIC-Net-Based LoS Positioning (Stage~2)}}
		\label{alg:musicnet_inference}
		\revision{\textbf{Input:} an SCM $\widehat{\boldsymbol R}$ and the estimated source number $\widehat K$ returned by Stage~1;}

		Feed $\widehat{\boldsymbol R}$ into MUSIC-Net to obtain the  complex-valued feature matrix $\boldsymbol X_{\!\boldsymbol W_{\!1}}$ and surrogate distance $\widehat r_0$\;
		
		Obtain the estimated noise subspace $\widehat{\boldsymbol U}_{\!n}$ 
		from $\boldsymbol X_{\!\boldsymbol W_{\!1}}$ using $\widehat K$ according to \eqref{eq:evd_of_surrogate}\;
		
		Apply the two-stage MUSIC procedure with $\widehat K$, $\widehat{\boldsymbol U}_{\!n}$, and $\widehat r_0$ to obtain 
		$\{(\widehat\theta_{k0},\widehat r_{k0})\}_{k=1}^{\widehat K}$\;

		\textbf{Output:} the estimated $\widehat K$ positions
		\revision{$\{\widehat{\boldsymbol p}_{k}\}_{k=1}^{\widehat K}$}, where
		$\widehat{\boldsymbol p}_{k}
		=
		\widehat r_{k0}
		[\cos(\widehat\theta_{k0}),\sin(\widehat\theta_{k0})]^T$.
	\end{algorithm}
	
	\revision{At inference, we use the estimated source number $\widehat K$ returned by Stage~1 in Section~\ref{sec:reliable_enumeration}.} Given \(\widehat K\), MUSIC-Net maps \(\widehat{\boldsymbol R}\) to the feature matrix \(\boldsymbol X_{\boldsymbol W_1}\) and surrogate distance \(\widehat r_0\). The noise subspace \(\widehat{\boldsymbol U}_n\) is then obtained from \(\boldsymbol{X}_{\boldsymbol W_1}\) via the EVD in \eqref{eq:evd_of_surrogate}, with the subspace dimension determined by \(\widehat K\). The estimated noise subspace $\widehat{\mv U}_{\!n}$ and surrogate distance $\widehat r_0$ are then fed into the two-stage MUSIC procedure to obtain the DoA estimates $\{\widehat\theta_{k0}\}_{k=1}^{\widehat K}$ and distance estimates $\{\widehat r_{k0}\}_{k=1}^{\widehat K}$, resulting in the estimated positions
	$\widehat{\boldsymbol p}_{k}=\widehat r_{k0}[\cos(\widehat\theta_{k0}),\sin(\widehat\theta_{k0})]^T$ for all detected sources $k\in[\widehat K]$. The above MUSIC-Net based positioning inference procedure is summarized in Algorithm~\ref{alg:musicnet_inference}.

	\section{\revision{Stage 3: SCP-Based Positioning Calibration}}
	\label{sec:conformal_calibration}
	\revision{Stage~2 provides the point position estimates for a test SCM $\widehat{\boldsymbol R}$. Using SCP, we construct an indexed collection $\widehat{\mathcal C}_p(\widehat{\boldsymbol R})$ of predicted regions to meet the joint coverage requirement in \eqref{eq:joint_position_coverage_requirement}. The calibration scores account for unordered position estimates and errors in source enumeration.}
	
	\subsection{Preliminary: Split Conformal Prediction \cite{angelopoulos2023conformal}}
	\label{subsec:SCP}
	As a special case of CRC, CP controls the miscoverage loss by constructing prediction sets with finite-sample marginal coverage guarantees \cite{angelopoulos_conformal_2024}. Among the various CP methods, SCP features simple implementation and computational efficiency, making it particularly suitable for finite-sample miscoverage control \cite{angelopoulos2023conformal}.  Let \(\mathcal D_{\mathrm{cal}}=\{(X^{(i)},Y^{(i)})\}_{i=1}^{n}\) be a calibration set. Given  a fixed predictive model and a  \emph{nonconformity score} function $s: \mathcal{X} \times \mathcal{Y} \rightarrow \mathbb{R}$, define the calibration score for each data point  $(X^{(i)}, Y^{(i)})$ by $S^{(i)}=s\left(X^{(i)}, Y^{(i)}\right)$, which measures how poorly the  $Y^{(i)}$ conforms to the  $X^{(i)}$. Then, for a target miscoverage level $\alpha \in(0,1)$ and a test feature $X^{(n+1)}$, SCP constructs a prediction set by 
	\begin{equation}
		\mathcal C(X^{(n+1)})
		= \{y \in\mathcal{Y} :s(X^{(n+1)},y)\le \widehat q_{1-\alpha}\},
		\label{eq:CP_set_generation}
	\end{equation}
	which contains candidate labels whose scores do not exceed the empirical quantile $\widehat q_{1-\alpha}$ given by
	\begin{equation}
		\widehat q_{1-\alpha}
		=
		Q\!\left(
		(1-\alpha)\left(1+\Myfrac{1}{n}\right);
		\{S^{(i)}\}_{i=1}^{n}
		\right),
		\label{eq:quantile_compute}
	\end{equation}
	where \(Q(\gamma;\{S^{(i)}\}_{i=1}^{n})\) is the \(\lceil n\gamma \rceil\)-th smallest value in the set  \(\{S^{(i)}\}_{i=1}^{n}\),  with the convention $Q(\gamma;\cdot)=+\infty$ when $\lceil n\gamma\rceil>n$.
	
	Under the assumption of exchangeable $\mathcal{D}_{\text{cal}}$ and $(X^{(n+1)}, Y^{(n+1)})$, the CP set \eqref{eq:CP_set_generation} satisfies the marginal coverage guarantee $\mathbb P\!\left( Y^{(n+1)} \in \mathcal C({X}^{(n+1)} )\right)\ge 1-\alpha$~\cite{angelopoulos2023conformal}.

	\subsection{Uncertainty Quantification for MUSIC-Net}
	\label{subsec:musicnet_uq}
	
	We provide in the sequel uncertainty quantification for positions of sources leveraging SCP. Define the calibration dataset for the positioning task as 
	$\mathcal{D}^{\text{po}}_{\text{cal}}
	=\{(\widehat{\boldsymbol{R}}^{(i)},
	\boldsymbol{\Theta}_{K^{(i)}})\}_{i=1}^{n}$. We assume that the calibration data samples in $\mathcal D_{\mathrm{cal}}^{\mathrm{po}}$ and a future test sample are exchangeable at the multi-source SCM level.
	
	Define
	$\mathfrak I_i=\mathfrak I(K^{(i)},\widehat K^{(i)})$.
	If $\mathfrak I_i\neq\emptyset$, we assume that there exists a deterministic matching scheme
	$\pi_i^\star\in\mathfrak I_i$, which can be obtained by solving a minimum-cost assignment problem via, e.g.,  the Hungarian algorithm~\cite{kuhn1955hungarian}. For the $i$-th calibration data sample, we define the nonconformity score function for DoA as
	\begin{align}
		\notag	S_\theta^{(i)}&=
		s_\theta\left(
		\widehat{\boldsymbol{R}}^{(i)},
		\{\theta_{k0}^{(i)}\}_{k=1}^{K^{(i)}}
		\right)
		\\&=
		\begin{cases}
			\displaystyle
			\max_{k\in[K^{(i)}]}
			\left|
			\theta_{k0}^{(i)}
			-
			\widehat\theta_{\pi_i^\star(k),0}^{(i)}
			\right|,
			& \widehat K^{(i)}\geq K^{(i)},\\[1ex]
			+\infty,
			& \widehat K^{(i)}<K^{(i)},
		\end{cases}
		\label{eq:theta_score_matched}
	\end{align}
	and that for distance as
	\begin{align}
		\notag	S_r^{(i)}
		&=
		s_r\left(
		\widehat{\boldsymbol{R}}^{(i)},
		\{r_{k0}^{(i)}\}_{k=1}^{K^{(i)}}
		\right)
		\\&=
		\begin{cases}
			\displaystyle
			\max_{k\in[K^{(i)}]}
			\left|
			r_{k0}^{(i)}
			-
			\widehat r_{\pi_i^\star(k),0}^{(i)}
			\right|,
			& \widehat K^{(i)}\geq K^{(i)},\\[1ex]
			+\infty,
			& \widehat K^{(i)}<K^{(i)}.
		\end{cases}
		\label{eq:range_score_matched}
	\end{align}
	Here, $\widehat K^{(i)}$ is obtained by the source enumeration method introduced in Section~\ref{sec:reliable_enumeration}, whereas $\widehat{\theta}_{k0}^{(i)}$ and $\widehat r_{k0}^{(i)}$ are the DoA and distance estimates produced by MUSIC-Net in Section~\ref{sec:subs_NN} for source $k\in[\widehat K^{(i)}]$.

	
	\begin{figure*}[!t]
		\centering
		\begin{minipage}[t]{0.48\textwidth}
			\centering
			\includegraphics[width=3.5in]{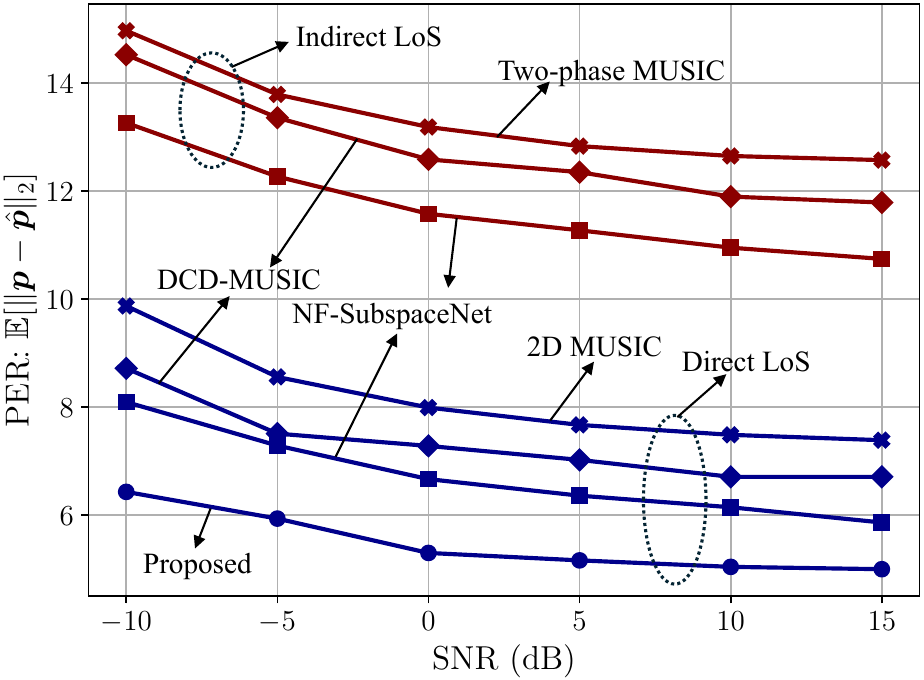}
			\caption{The mean positioning error (MPER) versus the SNR with $T=150$.}
			\label{fig:PER_SNR}
		\end{minipage}
		\hfill
		\begin{minipage}[t]{0.48\textwidth}
			\centering
			\includegraphics[width=3.5in]{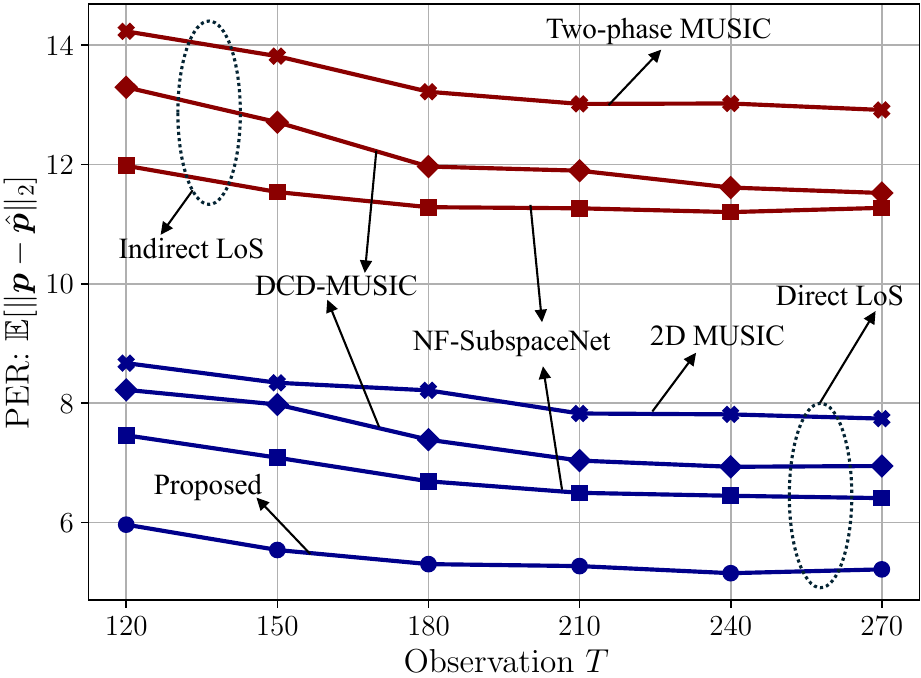}
			\caption{The MPER versus the number of observations $T$ with SNR $=5$ dB.}
			\label{fig:PER_T}
		\end{minipage}
	\end{figure*}
	
	%
	%

	Consider a test SCM $\widehat{\boldsymbol R}^{(n+1)}$ containing $K^{(n+1)}$ sources. Our goal is to construct an indexed collection of position predicted regions that satisfies the joint coverage requirement in \eqref{eq:joint_position_coverage_requirement}. Let $\widehat q_{\theta,1-\alpha_\theta}$ and $\widehat q_{r,1-\alpha_r}$ be the empirical quantiles obtained from \eqref{eq:quantile_compute} using target miscoverage levels $\alpha_\theta\in(0,1]$ and $\alpha_r\in(0,1]$, respectively. Write $\widehat K^{(n+1)}=\widehat K_{\widehat\eta}(\widehat{\boldsymbol R}^{(n+1)})$ for the Stage~1 source-number estimate and $\widehat K_{\mathrm{final}}^{(n+1)}$ for the final estimate after Stage~3. The latter is the source-number estimate used in the joint coverage requirement \eqref{eq:joint_position_coverage_requirement}.

	
	For each estimated source $k\in[\widehat K^{(n+1)}]$, the DoA prediction interval is
	\begin{equation}
		\mathcal C_{\theta,k}(\widehat{\boldsymbol R}^{(n+1)})
		=
		\left\{
		\theta \in \left[-\Myfrac{\pi}{2}, \Myfrac{\pi}{2}\right]
		~\middle|~
		\left|
		\theta-\widehat{\theta}_{k0}^{(n+1)}
		\right|
		\le
		\widehat q_{\theta,1-\alpha_\theta}
		\right\},
		\label{eq:CP_theta_source_set_construction}
	\end{equation}
	and the corresponding distance prediction interval is
	\begin{equation}
		\mathcal C_{r,k}(\widehat{\boldsymbol R}^{(n+1)})
		=
		\left\{
		r \in [r_{\min}, r_{\max}]
		~\middle|~
		\left|
		r-\widehat r_{k0}^{(n+1)}
		\right|
		\le
		\widehat q_{r,1-\alpha_r}
		\right\}.
		\label{eq:CP_r_source_set_construction}
	\end{equation}
	Combining \eqref{eq:CP_theta_source_set_construction} and
	\eqref{eq:CP_r_source_set_construction} yields the Cartesian position prediction region for estimated source $k$:
	\begin{equation}
		\begin{aligned}
			\mathcal C_{p,k}(\widehat{\boldsymbol R}^{(n+1)})
			&= \bigl\{ r[\cos\theta,\sin\theta]^T \,\bigm| \\
			&\quad \theta\in\mathcal C_{\theta,k}(\widehat{\boldsymbol R}^{(n+1)}), 
			r\in\mathcal C_{r,k}(\widehat{\boldsymbol R}^{(n+1)}) \bigr\}.
		\end{aligned}
		\label{eq:CP_cartesian_set_construction}
	\end{equation}

	{When both empirical quantiles
		$\widehat q_{\theta,1-\alpha_\theta}$ and
		$\widehat q_{r,1-\alpha_r}$ are finite, we set $\widehat K_{\mathrm{final}}^{(n+1)}=\widehat K^{(n+1)}$ and return the predicted regions
		constructed in \eqref{eq:CP_cartesian_set_construction} for all
		estimated sources. Otherwise, if either empirical quantile is infinite,
		we set $\widehat K_{\mathrm{final}}^{(n+1)}=K_{\max}\triangleq\tilde M-1$ and return
		$K_{\max}$ predicted regions, setting $\mathcal C_{p,j}(\widehat{\boldsymbol R}^{(n+1)})=\mathcal P_{\mathrm{all}}$ for every $j\in[K_{\max}]$, where the entire feasible position region is}
	\begin{equation}
		\mathcal P_{\mathrm{all}}
		=
		\left\{
		r[\cos\theta,\sin\theta]^T
		~\middle|~
		\theta\in
		\left[-\Myfrac{\pi}{2},\Myfrac{\pi}{2}\right],
		~
		r\in[r_{\min},r_{\max}]
		\right\}.
		\label{eq:entire_position_region}
	\end{equation} 
	Thus, the final indexed collection is
	\begin{equation}
		\begin{aligned}
			&\widehat{\mathcal C}_{p}(\widehat{\boldsymbol R}^{(n+1)}) \\
			&=
			\begin{cases}
				\displaystyle
				\left\{
				\mathcal C_{p,k}(\widehat{\boldsymbol R}^{(n+1)})
				\right\}_{k=1}^{\widehat K^{(n+1)}},
				&
				\begin{gathered}
					\widehat q_{\theta,1-\alpha_\theta}<+\infty \\
					\text{and}\quad
					\widehat q_{r,1-\alpha_r}<+\infty,
				\end{gathered}
				\\[1ex]
				\displaystyle
				\left\{
				\mathcal P_{\mathrm{all}}
				\right\}_{k=1}^{K_{\max}},
				&
				\text{otherwise}.
			\end{cases}
		\end{aligned}
		\label{eq:CP_output_collection}
	\end{equation}

	\revision{Since the source-number estimate is retained or increased, the MDR loss in \eqref{eq:source_under_loss} cannot increase.}
	The following proposition establishes the joint coverage guarantee for $\widehat{\mathcal C}_p(\widehat{\boldsymbol R}^{(n+1)})$.
	
	\begin{proposition}[Coverage Guarantee of the predicted regions]
		\label{prop:cartesian_coverage}
		Assume that the calibration data samples in
		$\mathcal{D}^{\text{po}}_{\text{cal}}
		=
		\{(\widehat{\boldsymbol{R}}^{(i)},
		\boldsymbol{\Theta}_{K^{(i)}}\}_{i=1}^{n}$
		and a test sample
		$(\widehat{\boldsymbol R}^{(n+1)},
		\boldsymbol{\Theta}_{K^{(n+1)}})$
		are exchangeable at the multi-source SCM level. Then,
		$\widehat{\mathcal C}_p(\widehat{\boldsymbol R}^{(n+1)})$
		defined in \eqref{eq:CP_output_collection} satisfies the joint position coverage requirement in
		\eqref{eq:joint_position_coverage_requirement},
		with $\alpha_p=\alpha_\theta+\alpha_r$.
	\end{proposition}
	
	\begin{IEEEproof}
		The proof is provided in Appendix~\ref{app:proof_cartesian_coverage}.
	\end{IEEEproof}

	\begin{figure*}[th]
		\centering
		
		\subfigure[$\alpha_e=0.1.$]{
			\includegraphics[height=0.391\textheight]{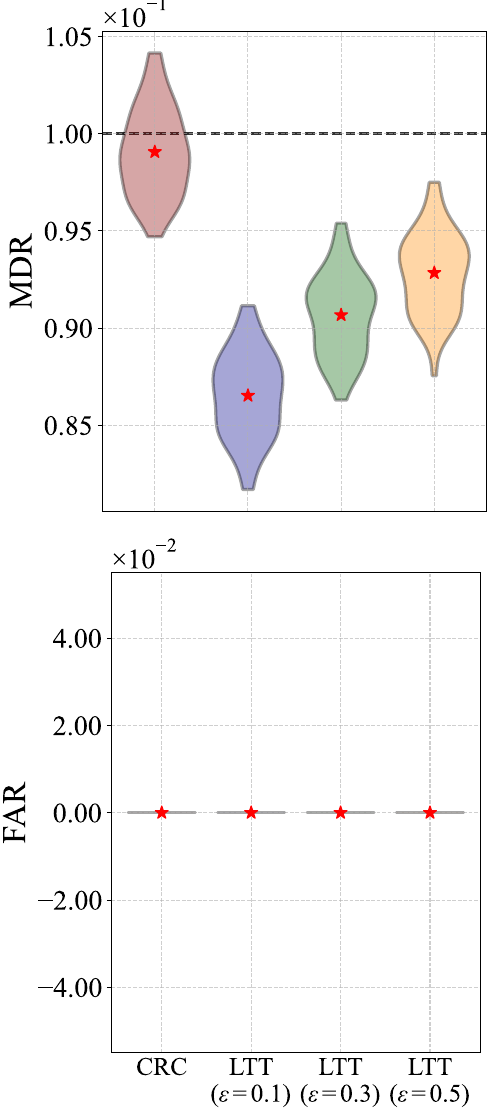}
			\label{fig:box07}
		}
		\subfigure[$\alpha_e=0.05.$]{
			\includegraphics[height=0.391\textheight]{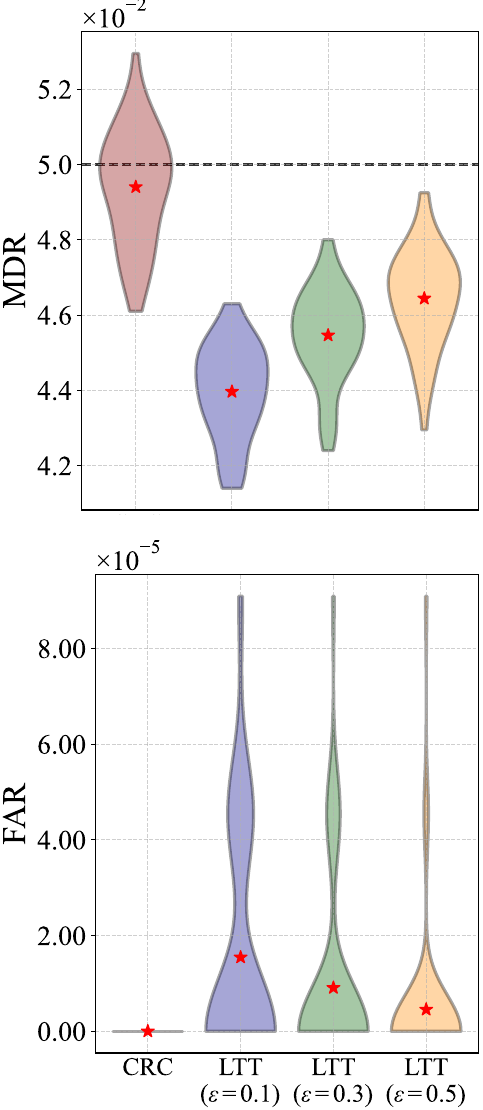}
			\label{fig:box08}
		}
		\subfigure[$\alpha_e=0.01.$]{
			\includegraphics[height=0.391\textheight]{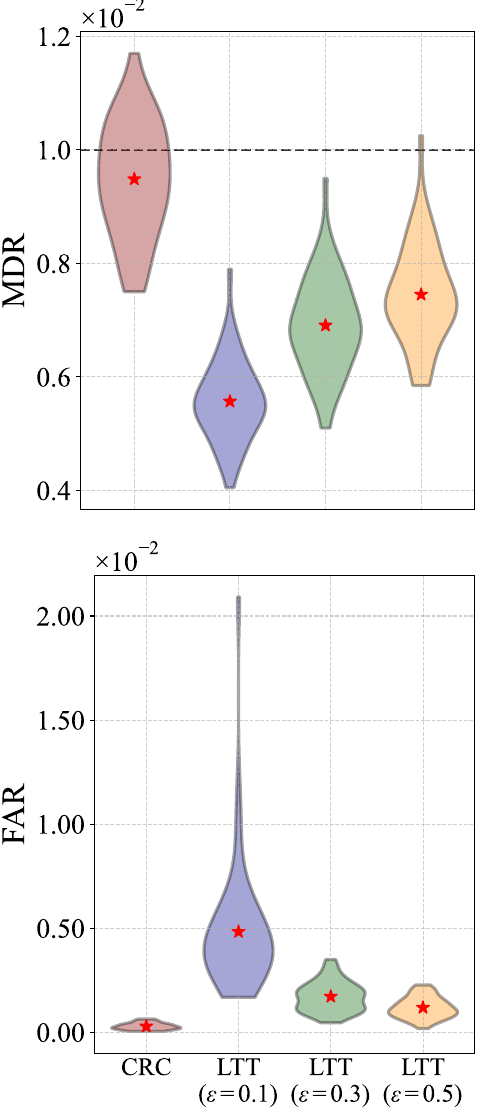}
			\label{fig:box09}
		}
		\subfigure[AIC and MDL baselines.]{
			\includegraphics[height=0.391\textheight]{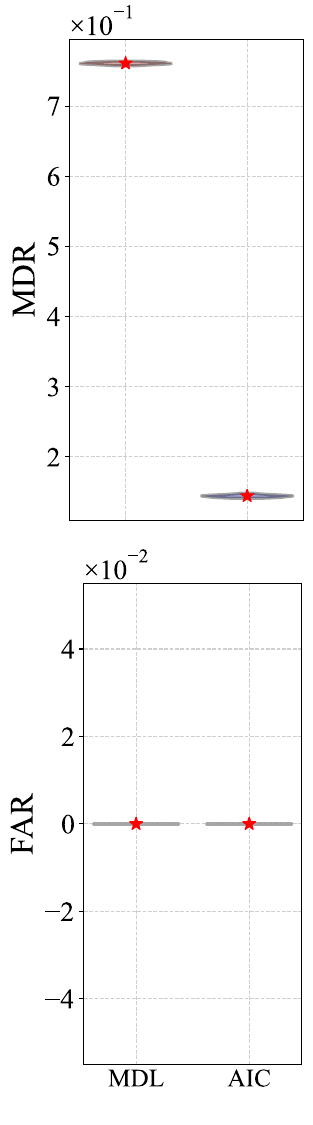}
			\label{fig:box10}
		}
		
		\vspace{-0.1in}
		\caption{Violin plots of the empirical MDR and FAR for different source enumeration schemes with $\mathrm{SNR}=-3$ dB and $T=150$. ``\textcolor{red}{$\star$}'' marks the empirical mean over $50$ Monte-Carlo trials.}
		\label{fig:box_all}
	\end{figure*}
	
	\begin{figure*}[t]
		\centering
		\subfigure[MDR versus SNR.]{
			\includegraphics[width=0.45\textwidth]{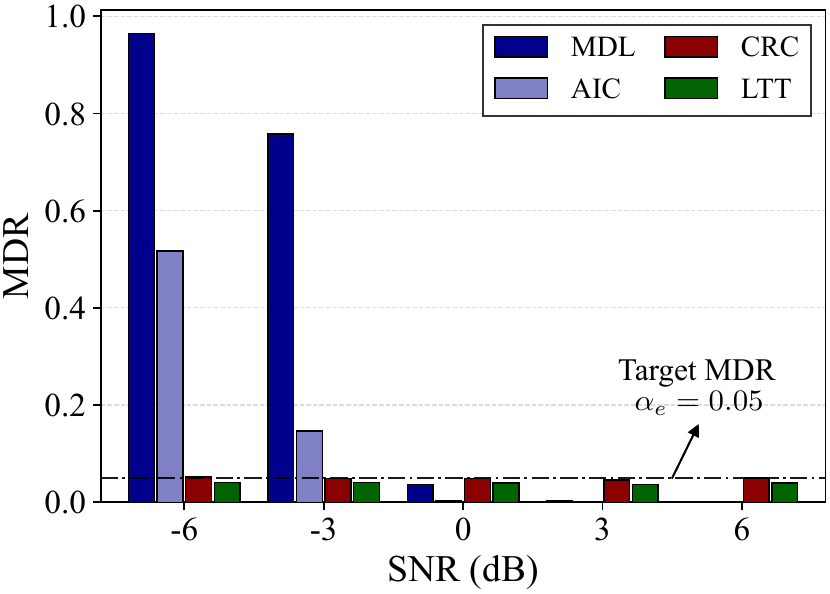}
			\label{fig:MDR_SNR}
		}
		\hfill
		\subfigure[FAR versus SNR.]{
			\includegraphics[width=0.45\textwidth]{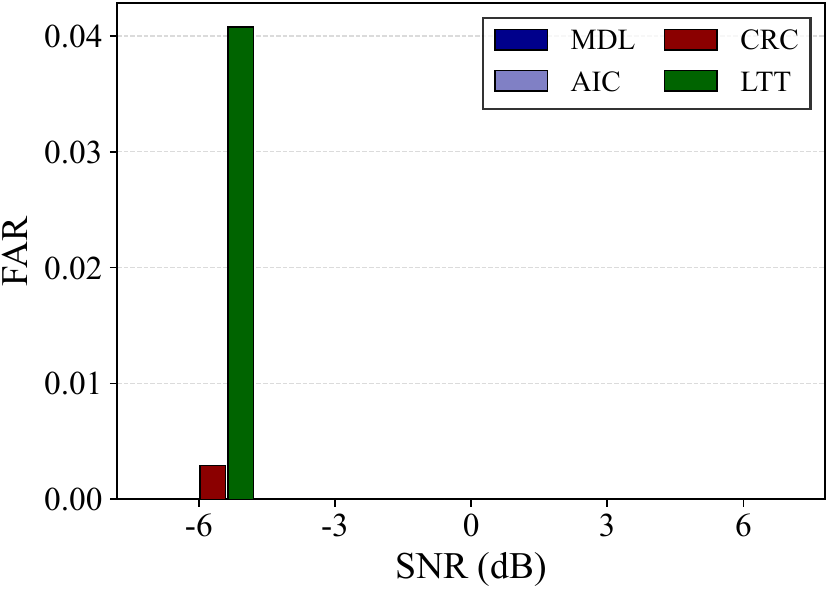}
			\label{fig:FAR_SNR}
		}
		\caption{Reliable source enumeration performance versus SNR with $T=150$, $\alpha_e=0.05$ and $\epsilon=0.5$.}
		\label{fig:enum_snr}
	\end{figure*}
	
	\section{Numerical Results}
	\label{sec:numerical_results}
	This section evaluates the proposed reliable near-field multi-source localization framework in a coherent-multipath setting.
	We assess: (i) the localization accuracy of MUSIC-Net, (ii) the MDR behavior of the CRC-calibrated source-enumeration scheme, and (iii) the coverage and efficiency of the SCP-based position predicted regions.

	\subsection{Experimental Settings}
	The experiment considers a near-field multi-source localization problem via uplink transmissions with a ULA of $\tilde M=129$ antennas operating at carrier frequency $f=30$\;GHz with half-wavelength antenna spacing. The number of sources is uniformly drawn from $K\in\{2,4,6,8\}$. For each source $k\in [K]$,  the DoA is sampled as $\theta_{k0}\sim\mathcal U(-60^\circ,60^\circ)$ subject to $\min_{i\neq j}|\theta_{i0}-\theta_{j0}|\geq4^\circ$. The distance is sampled as $r_{k0}^2\sim\mathcal U(10^2,50^2)$ m$^2$. Together, these ranges define the fan-shaped region considered in the experiments. For the NLoS paths $l\in[L_k]$, $k\in[K]$, the number $L_k$ of NLoS paths associated with source $k$ is drawn uniformly from $\{0,1,2,3\}$; the DoA $\theta_{kl}$ is generated by $\theta_{kl}\sim \mathcal U(\theta_{k0}-3^\circ,\theta_{k0}+3^\circ)$ with a minimum angular separation of at least $0.5^\circ$; the corresponding distance $r_{kl}$ is generated by $r_{kl}^2\sim \mathcal U((r_{k0}-2)^2,(r_{k0}+2)^2)$; and the  complex reflection coefficient is generated as $c_{kl}=\rho_{kl}e^{j\phi_{kl}}$ with the phase following $\phi_{kl}\sim \mathcal U(0,2\pi)$ and the amplitude following $\rho_{kl}\sim \mathcal U(0,1)$, respectively. Based on the generated LoS and NLoS parameters, i.e., $\boldsymbol{\Theta}_{K}$ and $\boldsymbol{\Psi}_{K}$, and given a prescribed transmit SNR ($ 10\log_{10}({\sum_{k=1}^K P_k}/{(K\sigma^2)})$ with equal source powers ($P_1 = P_2 = \cdots = P_K$), we synthesize the SCM $\widehat{\boldsymbol R}$  according to \eqref{eq:SCM} using $T$ snapshots at the prescribed transmit SNR. We generate $2.1\times10^5$ samples $(\widehat{\boldsymbol R},\boldsymbol\Theta_K)$: $1.8\times10^5$ for training, $2\times10^4$ for testing, and $1\times10^4$ for calibration. 
	For CRC-based source enumeration, each labeled calibration or test sample
	$(\widehat{\boldsymbol R},\boldsymbol{\Theta}_{K})$,
	where
	$\boldsymbol{\Theta}_{K}
	=(\theta_{k0},r_{k0})_{k=1}^{K}$,
	directly yields the corresponding source-number label $K$ and is thus used as
	$(\widehat{\boldsymbol R},K)$.

	We employ ResNet-based networks~\cite{he2016deep} as the neural architectures for the LoS feature extractor and surrogate distance estimator. The real and imaginary parts of $\widehat{\boldsymbol R}$ are stacked to form a two-channel input tensor.
	Then a $3\times3$ convolutional layer is followed by $N$ residual stages to extract high-level features. Each residual stage consists of
	a shortcut branch with a $1\times1$ convolutional layer and a residual branch
	with $L$ residual blocks, where each residual block contains two
	layers, each consisting of a rectified linear unit (ReLU) followed by a
	$3\times3$ convolution. The extracted high-level feature is further passed
	through a ReLU activation, average pooling, and a fully connected layer. The networks are trained using the Adam optimizer~\cite{kingma:adam} with a mini-batch size of $128$.

	We first compare MUSIC-Net with four positioning baselines. \textbf{2D MUSIC} directly estimates the positions of sources via a two-dimension angle-distance spectral search. \textbf{Two-phase MUSIC}~\cite{cheng2025comprime} applies spatial smoothing to decorrelate coherent DoAs but does not explicitly decorrelate coherent ranges. \textbf{NF-SubspaceNet}~\cite{Gast2026NFsubspace} employs an autoencoder to learn a surrogate covariance matrix for subsequent 2D MUSIC-based localization. \textbf{DCD-MUSIC}~\cite{gast2025dcd} employs two autoencoders to learn surrogate covariance matrices for cascaded DoA and range estimation, and further trains the cascaded architecture using positioning errors. 
	
	{Two-phase MUSIC}, {NF-SubspaceNet}, and {DCD-MUSIC} estimate all parameters including both $\boldsymbol{\Theta}_{K}$ and $\boldsymbol{\Psi}_{K}$. To enable a fair comparison with the LoS-positioning objective considered in this paper, we introduce the following two evaluation approaches for these schemes: Under the Indirect-LoS strategy, all path-related parameters are first estimated and then grouped source-wise using the path association strategy in~\cite[Algorithm 1]{xie2017source}. The delay-and-sum algorithm~\cite{widrow2005adaptive} is subsequently applied to each group to select the LoS angle-distance pair. Under the Direct-LoS strategy, the original outputs are replaced with E2E predictions of the multi-source positions.

	We next compare the proposed CRC-calibrated source-enumeration scheme with the following enumeration benchmarks: \textbf{AIC} and \textbf{MDL}, which are conventional information criteria using fixed penalty coefficients $\eta=2$ and $\eta=\ln T$, respectively, in \eqref{eq:source_number_selection}; and \textbf{Learn-then-Test (LTT)}~\cite{angelopoulos2025learn}, which calibrates the penalty coefficient  $\eta$ via finite-sample hypothesis testing to certify MDR control with probability at least $1-\epsilon$:
	\begin{align}
		\mathbb P
		\left(
		\mathbb E[\ell_{\mathrm{MDR}}(\widehat K_{\widehat\eta_{\mathrm{LTT}}}(\widehat{\boldsymbol R}),K)]
		\le \alpha_e
		\right)
		\ge 1-\epsilon, \label{LTT}
	\end{align}
	where $\epsilon\in(0,1)$ denotes the allowable failure probability and $\widehat\eta_{\mathrm{LTT}}$ is the penalty coefficient selected by LTT.

	Finally, for SCP-based positioning calibration, we compare the following benchmark nonconformity score function designs: \textbf{PER-SCP}, which uses the  positioning error  as the score function, i.e., $S_p^{(i)}=\max_{k\in[K^{(i)}]}\|\boldsymbol p_k^{(i)}-\widehat{\boldsymbol p}_{\pi_i^\star(k)}^{(i)}\|_2$ if $\widehat K^{(i)}\ge K^{(i)}$ and $S_p^{(i)}=+\infty$ otherwise,  and yields a circular prediction region around each estimated position, i.e., $\mathcal C_{p,k}^{\mathrm{PER}}(\widehat{\boldsymbol R}^{(n+1)})=\{\boldsymbol p\in\mathbb R^2:\|\boldsymbol p-\widehat{\boldsymbol p}_{k}^{(n+1)}\|_2\le \widehat q_{p,1-\alpha_p}\}$ for each estimated source $k\in[\widehat K^{(n+1)}]$; and \textbf{Axis-SCP}, which uses separate scene-level errors along the $x$- and $y$-axis as the score functions, i.e., $S_x^{(i)}=\max_{k\in[K^{(i)}]}|x_k^{(i)}-\widehat x_{\pi_i^\star(k)}^{(i)}|$ and $S_y^{(i)}=\max_{k\in[K^{(i)}]}|y_k^{(i)}-\widehat y_{\pi_i^\star(k)}^{(i)}|$ if $\widehat K^{(i)}\ge K^{(i)}$, and $S_x^{(i)}=S_y^{(i)}=+\infty$ otherwise, where $x_k^{(i)}=r_{k0}^{(i)}\cos\theta_{k0}^{(i)}$, $y_k^{(i)}=r_{k0}^{(i)}\sin\theta_{k0}^{(i)}$, $\widehat x_k^{(i)}=\widehat r_{k0}^{(i)}\cos\widehat\theta_{k0}^{(i)}$, and $\widehat y_k^{(i)}=\widehat r_{k0}^{(i)}\sin\widehat\theta_{k0}^{(i)}$, thus yielding a rectangular prediction region for each estimated source $k\in[\widehat K^{(n+1)}]$, i.e., $\mathcal C_{p,k}^{\mathrm{Axis}}(\widehat{\boldsymbol R}^{(n+1)})=[\widehat x_k^{(n+1)}-\widehat q_{x,1-\alpha_x},\widehat x_k^{(n+1)}+\widehat q_{x,1-\alpha_x}]\times[\widehat y_k^{(n+1)}-\widehat q_{y,1-\alpha_y},\widehat y_k^{(n+1)}+\widehat q_{y,1-\alpha_y}]$, where $\alpha_x > 0 $ and $\alpha_y > 0$ are the target miscoverage levels for the $x$- and $y$-axis prediction intervals, respectively, satisfying $\alpha_x+\alpha_y=\alpha_p$.
	
	

	\subsection{Numerical Results}
	\subsubsection{Point Estimation Performance}
	We measure the point-estimation accuracy using the MPER between the 
	estimated position $\widehat{\boldsymbol p}$ and the ground truth 
	$\boldsymbol p$, defined as 
	$\mathbb{E}\|\boldsymbol p-\widehat{\boldsymbol p}\|_2$, where the expectation 
	is taken over all samples in the test dataset. Fig.~\ref{fig:PER_SNR} and Fig.~\ref{fig:PER_T} show that the MUSIC-Net achieves the lowest observed MPER across different SNR regimes and numbers of observations. Moreover, the direct-LoS variants generally outperform their indirect-LoS counterparts. This observation is consistent with the additional errors that can arise during path estimation and subsequent LoS-path selection when all propagation paths are estimated before the LoS component is extracted.

	\subsubsection{Reliable Source Enumeration Performance}
	We evaluate source enumeration using the test-set averages of the MDR in \eqref{eq:source_under_loss} and $\mathrm{FAR}=\max(\widehat K-K,0)/\max(1, \widehat K)$, which measures normalized source-number overestimation.


	We first validate the MDR control capability of the proposed CRC-calibrated source enumeration scheme, as required in \eqref{eq:reliable_source_enum_goal}. Fig.~\ref{fig:box_all} shows that the proposed scheme maintains the mean empirical MDR at or below the prescribed level $\alpha_e$, while keeping the FAR low. The LTT-based schemes yield lower MDR values as $\epsilon$ decreases, at the cost of larger FAR values under stricter MDR targets. In contrast, AIC and MDL are not calibrated to the prescribed MDR level and exhibit MDRs above all three target levels despite negligible FARs.

	\begin{figure}[h]
		\centering
		\includegraphics[width=\linewidth]{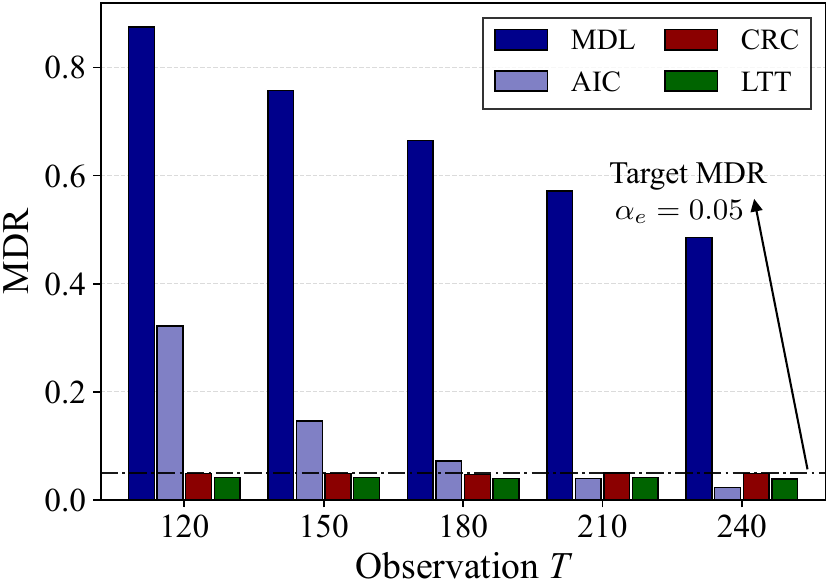}
		\caption{MDR versus the number of observations $T$ with
			$\mathrm{SNR}=-3$ dB, $\alpha_e=0.05$, and $\epsilon=0.5$.
			All four schemes yield zero empirical FAR over the tested range of $T$.}
		\label{fig:enum_T}
	\end{figure}
	\begin{figure}[h]
		\centering
		\includegraphics[width=\linewidth]{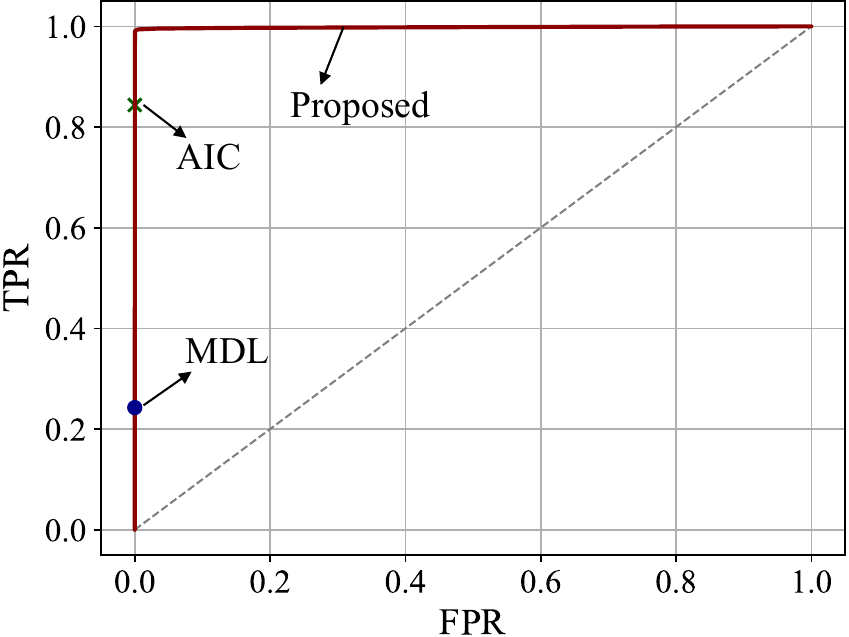}
		\caption{ROC comparison at $\mathrm{SNR}=-3$ dB and $T=150$.}
		\label{fig:ROC}
	\end{figure}

	\begin{figure*}[t]
		\noindent\hspace*{-0.02\textwidth}%
		\subfigure[$1-\alpha_p=0.7$.]{
			\includegraphics[width=0.331\textwidth]{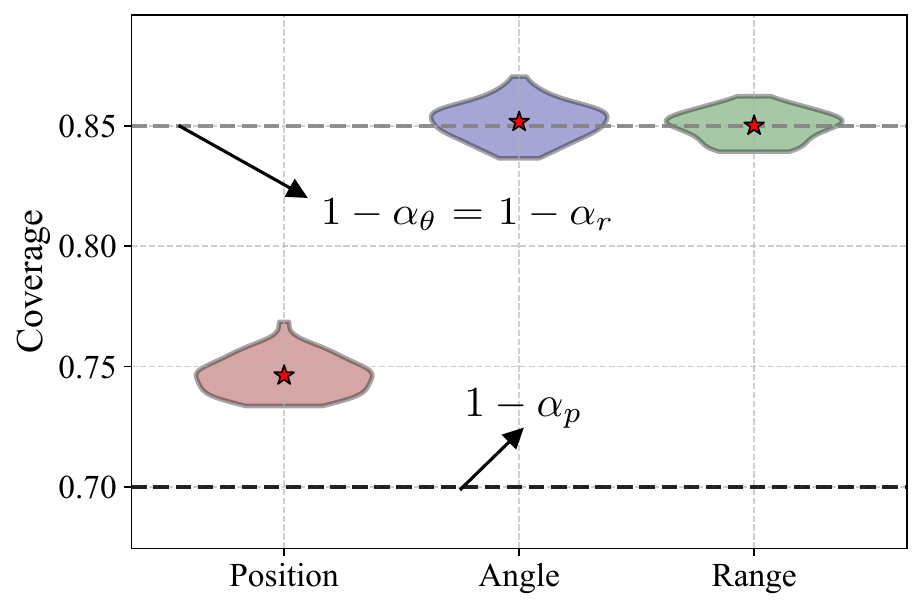}
			\label{fig:coverage_alpha_p_07}
		}
		\hspace{-0.02\textwidth}%
		\subfigure[$1-\alpha_p=0.8$.]{
			\includegraphics[width=0.331\textwidth]{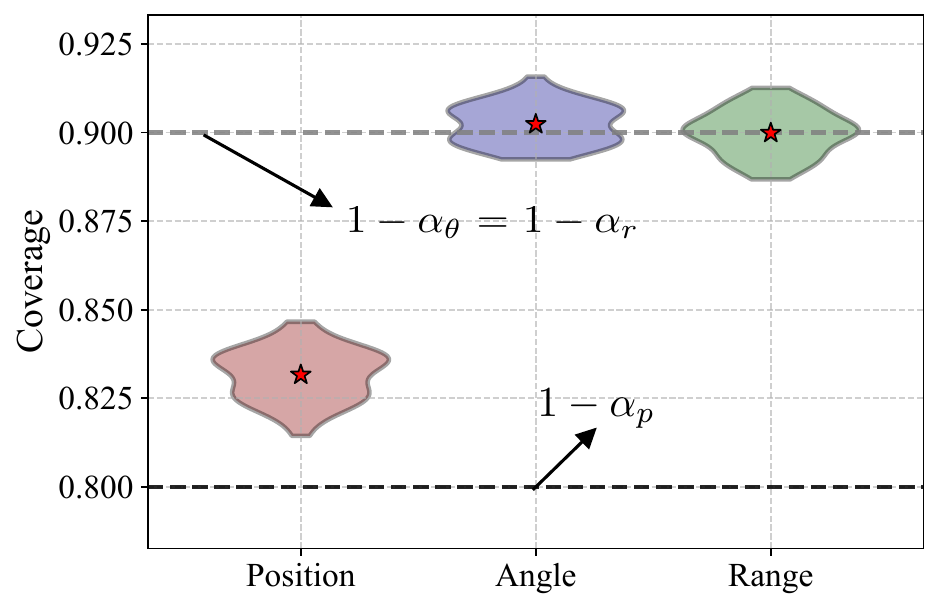}
			\label{fig:coverage_alpha_p_08}
		}
		\hspace{-0.02\textwidth}%
		\subfigure[$1-\alpha_p=0.9$.]{
			\includegraphics[width=0.331\textwidth]{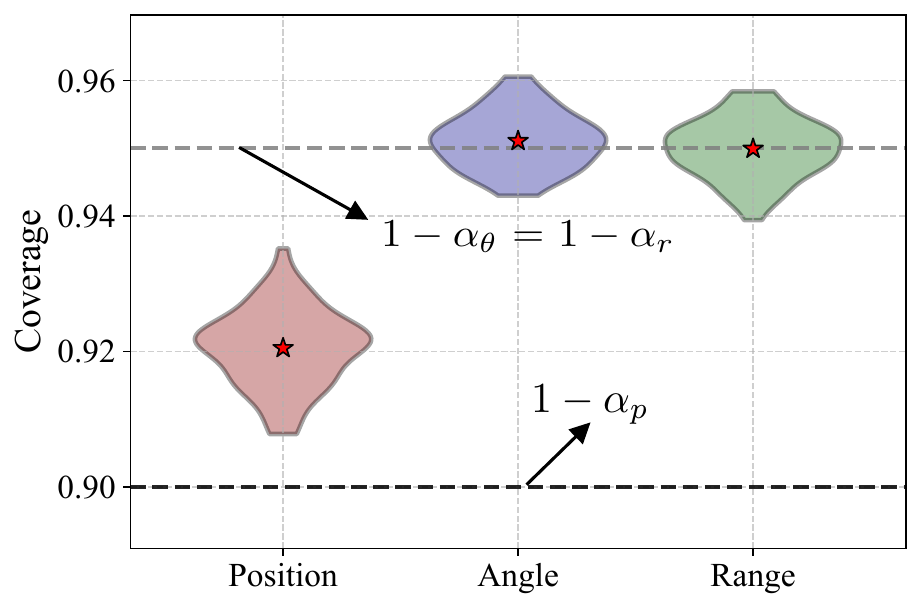}
			\label{fig:coverage_alpha_p_09}
		}
		\caption{Violin plots of the empirical coverage probability achieved under different target coverage probability $1-\alpha_p$ with $\mathrm{SNR}=5$ dB, $T=150$ and $\alpha_\theta = \alpha_r =\alpha_p/2 $. ``\textcolor{red}{$\star$}'' denotes the empirical mean over $50$ Monte-Carlo trials.}
		\label{fig:coverage_violin_results}
	\end{figure*}

	\begin{figure*}[t]
		
		\noindent\hspace*{-0.02\textwidth}%
		\subfigure[Position.]{
			\includegraphics[width=0.331\textwidth]{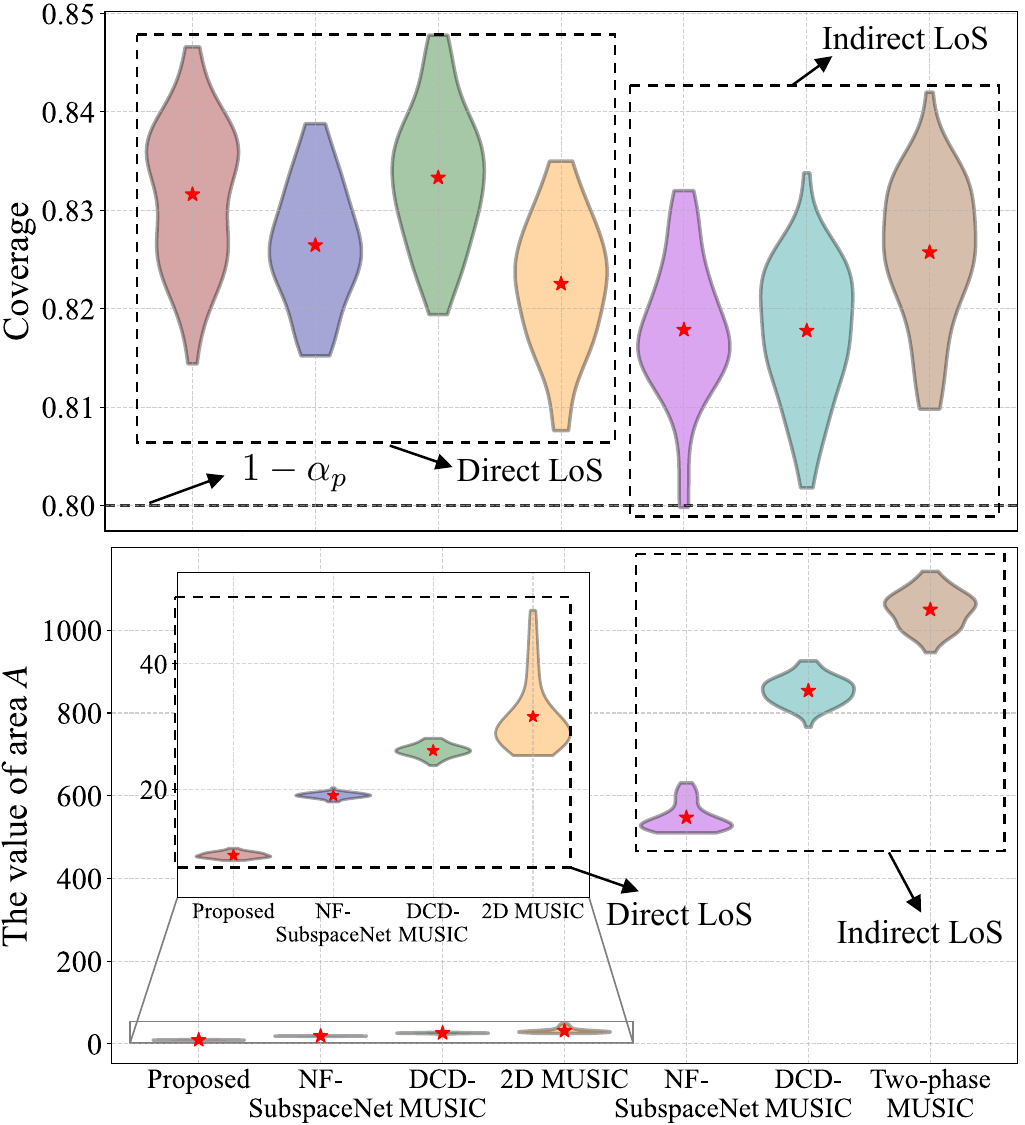}
			\label{fig:po_all_area_07}
		}
		\hspace{-0.02\textwidth}%
		\subfigure[DoA.]{
			\includegraphics[width=0.331\textwidth]{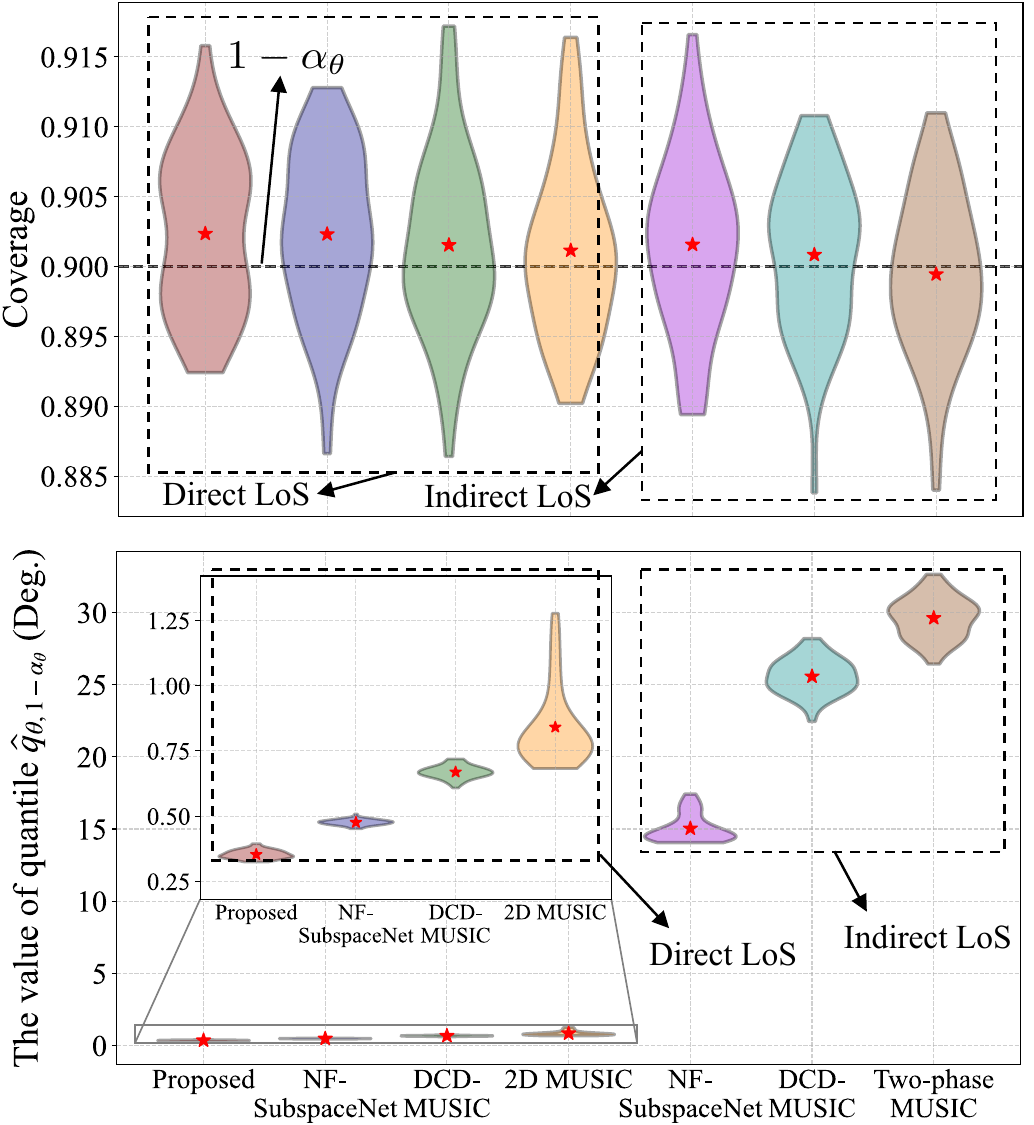}
			\label{fig:po_all_area_08}
		}
		\hspace{-0.02\textwidth}%
		\subfigure[Distance.]{
			\includegraphics[width=0.331\textwidth]{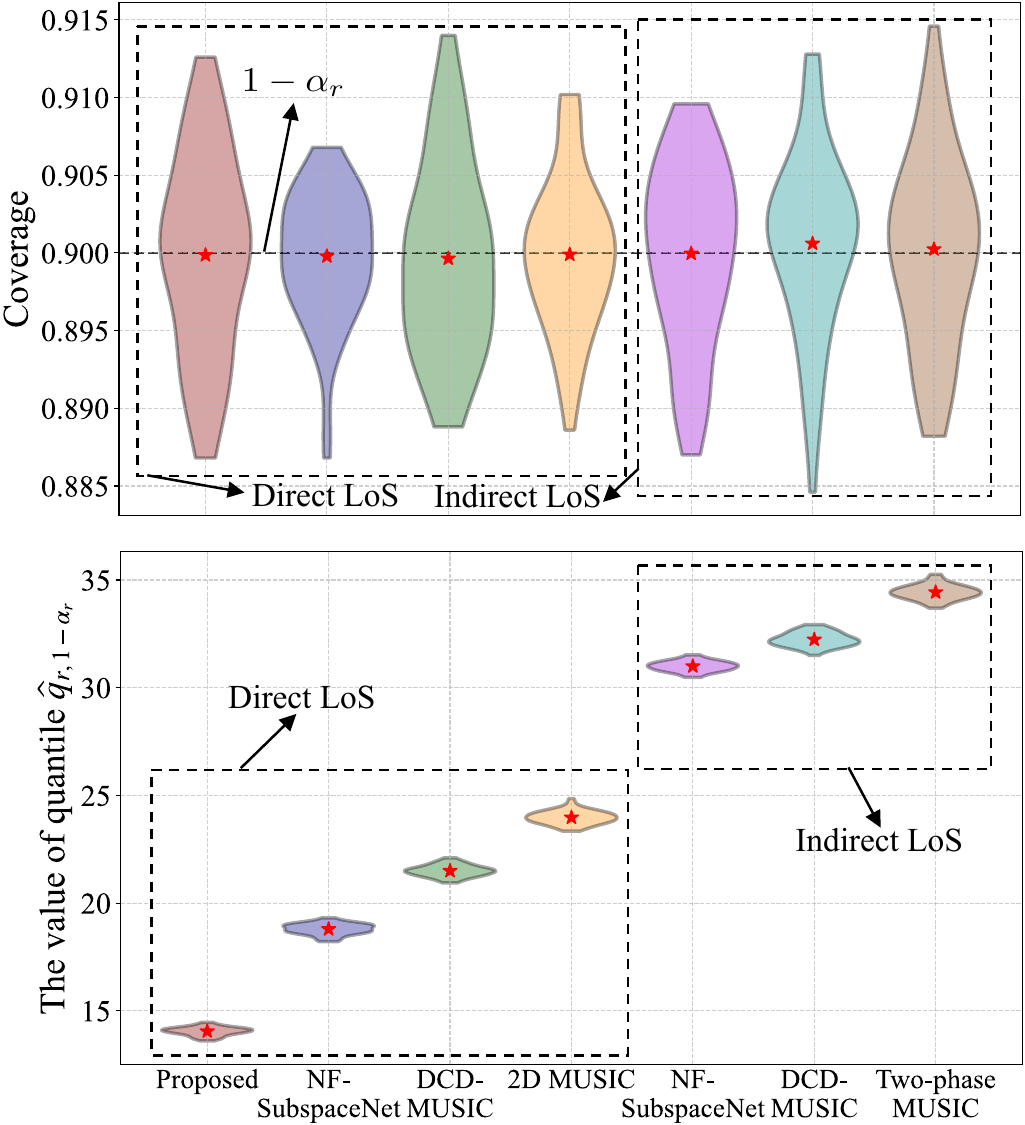}
			\label{fig:po_all_area_09}
		}
		
		\caption{Violin plots of the empirical coverage probabilities and corresponding efficiency metrics for position, DoA, and distance achieved by different positioning schemes
			with $\mathrm{SNR}=5$ dB, $T=150$, $\alpha_\theta=\alpha_r = \alpha_p/2 =0.1$. ``\textcolor{red}{$\star$}'' denotes the empirical mean over $50$ Monte-Carlo trials.}
		\label{fig:po_all}
	\end{figure*}

	Next, Figs.~\ref{fig:enum_snr} and~\ref{fig:enum_T} evaluate the robustness of the source enumeration schemes under different SNR regimes and different numbers of observations $T$, respectively. Fig.~\ref{fig:enum_snr} shows that the proposed CRC-based scheme maintains the empirical MDR at or below the prescribed level, while the LTT-based scheme yields slightly lower MDR values at the cost of a larger FAR at low SNR regime. As the SNR increases, the FARs of both schemes approach zero. In contrast, AIC and MDL exhibit negligible FARs but suffer from relatively large MDR values at low SNR regime. Their MDRs decrease as the SNR increases and satisfy the prescribed MDR level at higher SNRs. Similar trends are observed in Fig.~\ref{fig:enum_T}. The proposed scheme consistently satisfies the MDR target across the considered values of $T$. Increasing $T$ reduces the MDRs of AIC and MDL, with AIC meeting the target at larger tested values of $T$, whereas MDL remains above it. These results indicate that the proposed reliable scheme provides stable empirical MDR control, whereas conventional methods, including AIC and MDL, require higher SNRs or more observations to meet the prescribed MDR requirement.

	Finally, Fig.~\ref{fig:ROC} compares the Receiver operating characteristic (ROC) curve of the proposed scheme with the operating points of AIC and MDL, where the true positive rate (TPR) is given by $\mathrm{TPR}=1-\mathrm{MDR}$ and the false positive rate (FPR) is represented by the FAR. At comparable near-zero FAR values, the proposed scheme achieves a higher TPR and hence a lower MDR, indicating a more favorable tradeoff between source-number underestimation and overestimation under the considered setting.

	\subsubsection{Positioning Calibration Performance}


	\begin{figure*}[t]
		\noindent\hspace{-0.02\textwidth}
		\subfigure[Coverage probability.]{
			\includegraphics[width=0.311\textwidth]{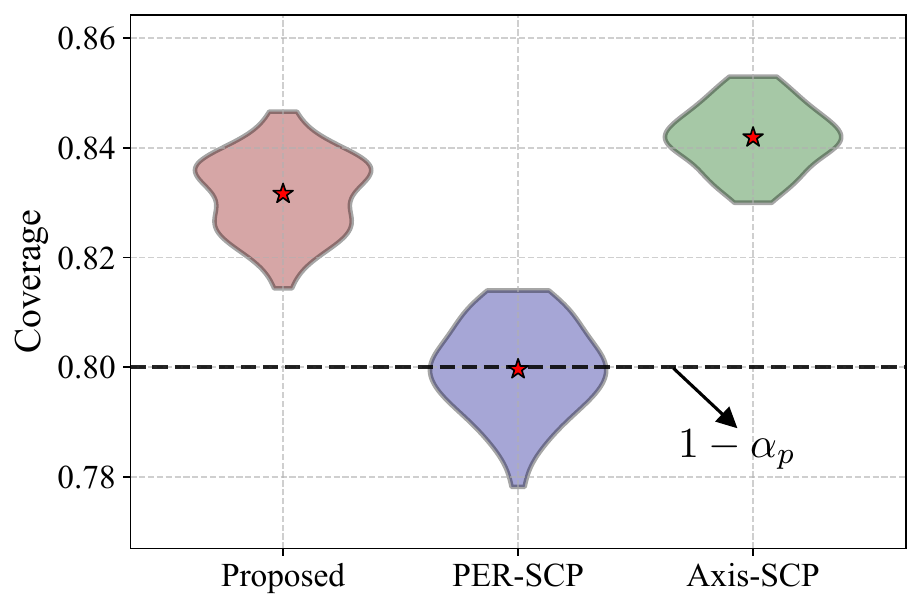}
			\label{fig:po_score_design_joint_coverage}
		}
		\hspace{-0.02\textwidth}%
		\subfigure[ Average set size of ${\mathcal C}_{p,k}(\widehat{\boldsymbol R}^{(n+1)})$.]{
			\includegraphics[width=0.311\textwidth]{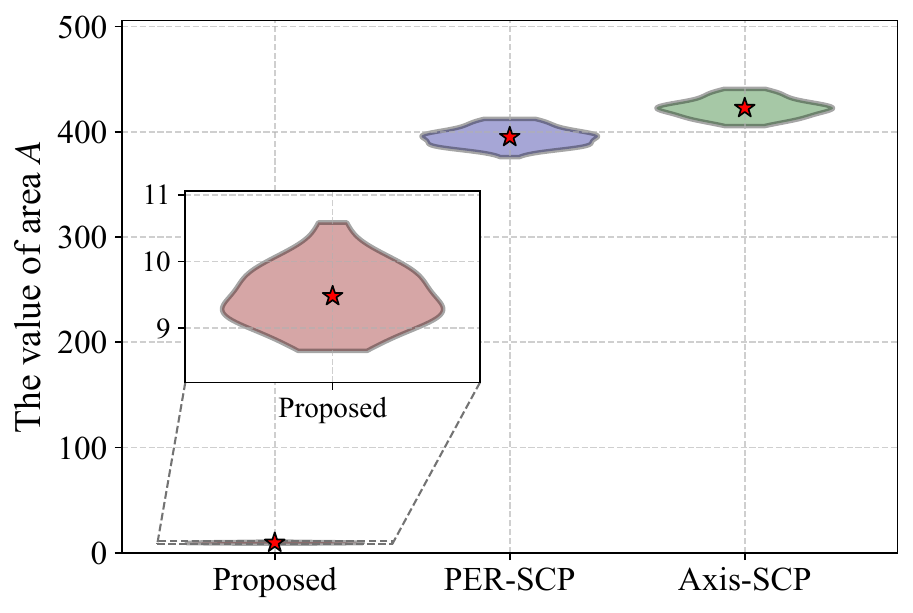}
			\label{fig:po_score_design_area}
		}
		\hspace{-0.01\textwidth}%
		\subfigure[ Visualization of ${\mathcal C}_{p,k}(\widehat{\boldsymbol R}^{(n+1)})$.]{
			\includegraphics[width=0.38\textwidth]{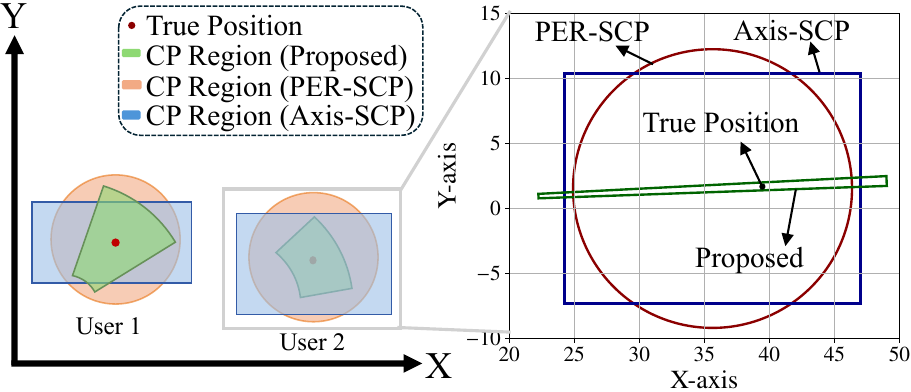}
			\label{fig:po_score_design_visualize}
		}
		
		\caption{Comparison of different SCP score function designs with target coverage level $1-\alpha_p=0.8$, $\mathrm{SNR}=5$ dB, and $T=150$. ``\textcolor{red}{$\star$}'' denotes the empirical mean over $50$ Monte-Carlo trials.}
		\label{fig:po_score_design}
	\end{figure*}

	\begin{figure*}[t]
		\noindent\hspace{-0.02\textwidth}
		\subfigure[Average set size of ${\mathcal C}_{p,k}(\widehat{\boldsymbol R}^{(n+1)})$.]{
			\includegraphics[width=0.341\textwidth]{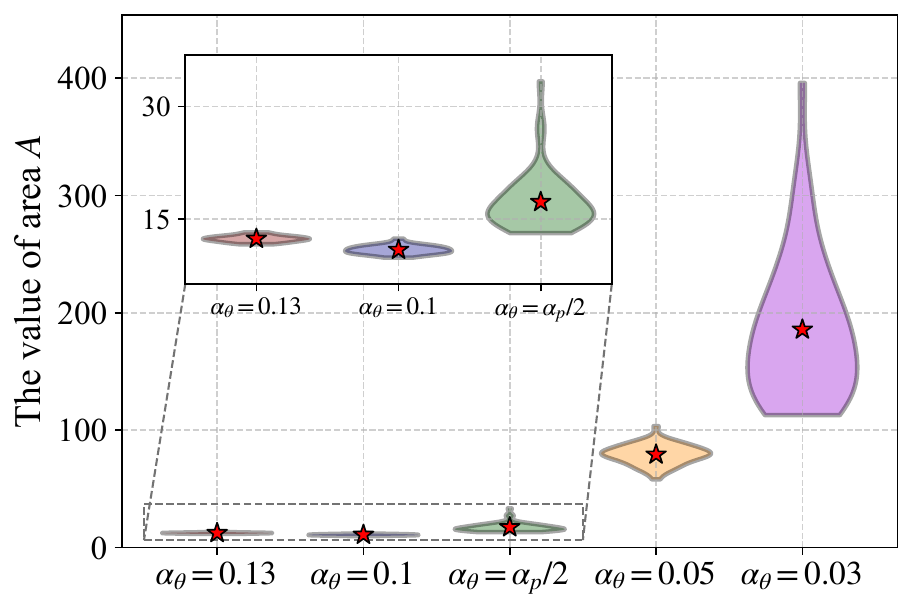}
			\label{fig:allocation_strategy_area}
		}
		\hspace{-0.02\textwidth}%
		\subfigure[ DoA quantile $\widehat{q}_{\theta,1-\alpha_\theta}$.]{
			\includegraphics[width=0.32\textwidth]{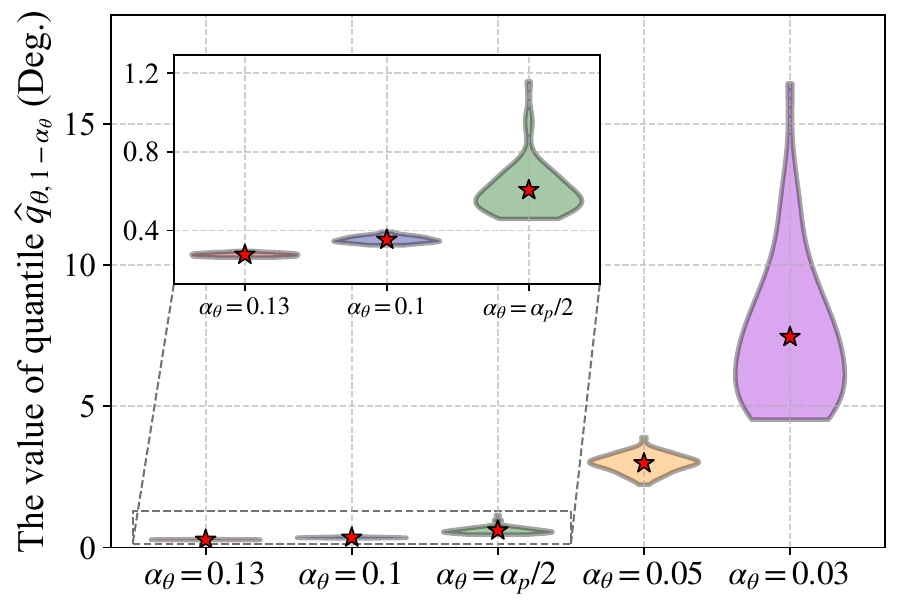}
			\label{fig:allocation_strategy_DoA}
		}
		\hspace{-0.02\textwidth}%
		\subfigure[Distance quantile $\widehat{q}_{r,1-\alpha_r}$.]{
			\includegraphics[width=0.32\textwidth]{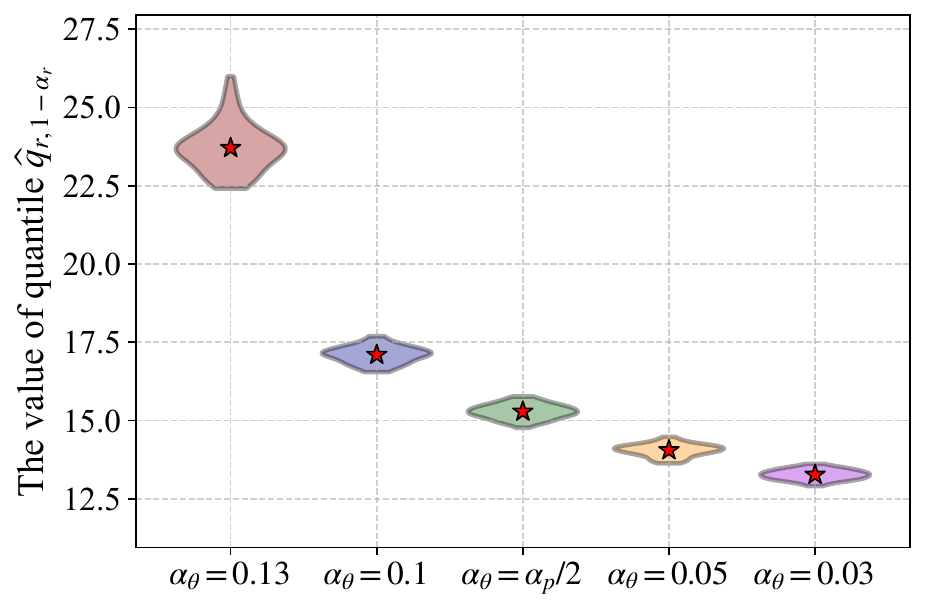}
			\label{fig:allocation_strategy_distance}
		} 
		\caption{Comparison of different miscoverage allocation strategies in terms of the average set size $A$ of ${\mathcal C}_{p,k}(\widehat{\boldsymbol R}^{(n+1)})$, the DoA quantile $\widehat q_{\theta,1-\alpha_\theta}$, and the distance quantile $\widehat q_{r,1-\alpha_r}$ with  $\mathrm{SNR}=5$ dB, $T=150$, $1-\alpha_p=0.85$ and $\alpha_p=\alpha_\theta+\alpha_r$. ``\textcolor{red}{$\star$}'' denotes the empirical mean over $50$ Monte-Carlo trials.}
		\label{fig:allocation_strategy}
	\end{figure*}

	We evaluate the  empirical coverage probability and the average area, denoted by $A$, of each prediction region $\mathcal{C}_{p,k}(\widehat{\boldsymbol R}^{(n+1)})$. For the proposed SCP scheme, $A=\mathbb E[4\widehat r_{k0}\widehat q_{r,1-\alpha_r}\widehat q_{\theta,1-\alpha_\theta}]$, where the expectation is taken over all estimated sources and test samples. For PER-SCP and Axis-SCP, the sizes of the region are given by $A=\pi\widehat q_{p,1-\alpha_p}^{2}$ and $A=4\widehat q_{x,1-\alpha_x}\widehat q_{y,1-\alpha_y}$, respectively.

	Fig.~\ref{fig:coverage_violin_results} evaluates coverage at different target levels.
	The empirical DoA and distance coverages closely follow their respective target levels $1-\alpha_\theta$ and $1-\alpha_r$, respectively. Their joint empirical  coverage exceeds the target level $1-\alpha_p$, which is consistent with the conservativeness introduced by combining the separate DoA and distance guarantees through the union bound in \eqref{eq:joint_score_coverage}.

	Fig.~\ref{fig:po_all} compares coverage and efficiency across the positioning methods. All SCP-calibrated methods satisfy \eqref{eq:joint_position_coverage_requirement}, while the proposed MUSIC-Net yields the smallest $\mathcal C_{p,k}(\widehat{\boldsymbol R}^{(n+1)})$ because of its lower point-estimation error.

	Figs.~\ref{fig:po_score_design_joint_coverage} and \ref{fig:po_score_design_area} compare the three nonconformity-score designs. All three designs satisfy the prescribed coverage level, but the proposed design yields the smallest prediction region area, and therefore the most efficient uncertainty quantification. The gain arises from  the geometry induced by the score functions. The proposed design calibrates the DoA and distance estimates before the polar-to-Cartesian transformation  and produces an annular sector, whereas PER-SCP and Axis-SCP construct circular and axis-aligned rectangular regions, respectively, as illustrated in 
	Fig.~\ref{fig:po_score_design_visualize}.

	Finally, Fig.~\ref{fig:allocation_strategy} examines the allocation of $\alpha_p=\alpha_\theta+\alpha_r$.
	The average area of the prediction region  $\mathcal{C}_{p,k}(\widehat{\boldsymbol R}^{(n+1)})$ decreases substantially as a larger fraction of $\alpha_p$ is assigned to DoA, but increases slightly as $\alpha_\theta$ approaches $\alpha_p$. Because near-field DoA estimates are typically more accurate than distance estimates \cite{liang_2026_RIS_near_field}, an asymmetric allocation with $\alpha_\theta>\alpha_r$ imposes a stricter coverage requirement on distance.  Although the smaller $\alpha_r$ increases $\widehat q_{r,1-\alpha_r}$, the resulting increase can be offset by the larger reduction in $\widehat q_{\theta,1-\alpha_\theta}$. This tradeoff provides a practical rule for improving distance reliability without unnecessarily enlarging the position prediction region.

	\section{Conclusion}
	\label{sec:conclusion}
	This paper developed a reliability-aware framework for near-field multi-source enumeration and positioning in coherent multipath environments. The framework calibrates a penalized-likelihood source-enumeration rule via CRC to control the expected MDR, and MUSIC-Net recovers the LoS feature subspace and surrogate distance required for direct two-stage-MUSIC positioning without estimating or associating NLoS paths. SCP then constructs an indexed collection of position predicted regions with a finite-sample joint coverage guarantee. Numerical results showed that MUSIC-Net achieved the lowest observed MPER among the considered methods, CRC kept the empirical MDR close to its prescribed level while incurring smaller FARs than LTT under comparable settings, and the proposed SCP design satisfied the target coverage with smaller predicted regions. These results demonstrate the value of treating source enumeration and positioning as sequential reliability-critical tasks.

	\appendix

	\subsection{Proof of Lemma~\ref{lem:source_loss_monotone}}
	\label{app:proof_source_loss_monotone}
	
	Fix an SCM $\widehat{\boldsymbol R}$ and write $\widehat K_\eta=\widehat K_\eta(\widehat{\boldsymbol R})$. We first show that $\widehat K_\eta$ can be chosen to be monotone non-increasing in $\eta$. 
	From \eqref{eq:source_number_selection}, define
	\begin{align}
		D(q)=-2\ln p(\boldsymbol Y;\widehat{\boldsymbol\Xi}_q,q).
	\end{align}
	and know that $G(q)$ can be directly verified that it is strictly increasing over 
	$q\in\{0,\ldots,\tilde M-1\}$, so that \eqref{eq:source_number_selection} can be simplified to 
	\begin{align}
		\widehat K_\eta\in\arg\min_q\left[D(q)+\eta G(q)\right].
	\end{align}
	Then, let
	\begin{equation}
		q_1\in \arg\min_{q}
		\left[D(q)+\eta_1 G(q)\right],
	\end{equation}
	and
	\begin{equation}
		q_2\in \arg\min_{q}
		\left[D(q)+\eta_2 G(q)\right],
	\end{equation}
	where $0\le \eta_1\le \eta_2$. 
	By the optimality of $q_1$ and $q_2$, we have
	\begin{equation}
		D(q_1)+\eta_1 G(q_1)
		\le
		D(q_2)+\eta_1 G(q_2),
		\label{eq:app1_eq1}
	\end{equation}
	and
	\begin{equation}
		D(q_2)+\eta_2 G(q_2)
		\le
		D(q_1)+\eta_2 G(q_1).
		\label{eq:app1_eq2}
	\end{equation}
	Rearranging \eqref{eq:app1_eq1} and \eqref{eq:app1_eq2} yields
	\begin{equation}
		D(q_1)-D(q_2)
		\le
		\eta_1\bigl(G(q_2)-G(q_1)\bigr),
		\label{eq:app1_eq3}
	\end{equation}
	and
	\begin{equation}
		D(q_1)-D(q_2)
		\ge
		\eta_2\bigl(G(q_2)-G(q_1)\bigr).
		\label{eq:app1_eq4}
	\end{equation}
	Combining \eqref{eq:app1_eq3} and \eqref{eq:app1_eq4}, we obtain
	\begin{equation}
		(\eta_2-\eta_1)
		\bigl(G(q_2)-G(q_1)\bigr)
		\le 0.
	\end{equation}
	Since $\eta_2-\eta_1\ge 0$, it follows that $G(q_2)\le G(q_1).$
	Because $G(q)$ is strictly increasing in $q \in \{ 0,\ldots,\tilde{M}-1 \}$, we conclude that
	\begin{align}
		q_2\le q_1
	\end{align}
	with $0 \le \eta_1 \le \eta_2$.
	Therefore, $\widehat K_\eta$ can be chosen to be monotone non-increasing in $\eta$.
	
	For a fixed true source number $K\ge 1$, the function $\max(K-x,0)$ is monotone non-increasing in $x$. Hence, since $\widehat K_\eta$ is monotone non-increasing in $\eta$, the quantity $\max(K-\widehat K_\eta,0)$
	is monotone non-decreasing in $\eta$. Thus, the MDR loss in \eqref{eq:source_under_loss} with $\widehat K_\eta$, i.e.,
	\[
	\ell_{\mathrm{MDR}}
	\left(
	\widehat K_\eta,K
	\right)
	=
	\frac{\max(K-\widehat K_\eta,0)}{K}
	\]
	is monotone non-decreasing in $\eta$. This completes the proof.
	\subsection{Proof of Proposition~\ref{prop:cartesian_coverage}}
	\label{app:proof_cartesian_coverage}
	
	Let $\widehat K_{\mathrm{final}}^{(n+1)}$ and $\widehat{\mathcal C}_p(\widehat{\boldsymbol R}^{(n+1)})$ denote the final source-number estimate and indexed collection of predicted regions, respectively. By the marginal coverage guarantee of SCP~\cite{angelopoulos2023conformal}, we have
	\begin{equation}
		\mathbb P\left(
		S_\theta^{(n+1)}
		\le
		\widehat q_{\theta,1-\alpha_\theta}
		\right)
		\ge
		1-\alpha_\theta,
		\label{eq:theta_score_coverage}
	\end{equation}
	and
	\begin{equation}
		\mathbb P\left(
		S_r^{(n+1)}
		\le
		\widehat q_{r,1-\alpha_r}
		\right)
		\ge
		1-\alpha_r.
		\label{eq:range_score_coverage}
	\end{equation}
	Therefore, by the union bound and
	$\alpha_p=\alpha_\theta+\alpha_r$,
	\begin{align}
		\mathbb P\left(
		S_\theta^{(n+1)}
		\le
		\widehat q_{\theta,1-\alpha_\theta},
		~
		S_r^{(n+1)}
		\le
		\widehat q_{r,1-\alpha_r}
		\right)
		\ge
		1-\alpha_p.
		\label{eq:joint_score_coverage}
	\end{align}
	
	Define the joint  coverage event
	\[
	\mathcal E
	=
	\left\{
	S_\theta^{(n+1)}
	\le
	\widehat q_{\theta,1-\alpha_\theta},
	~
	S_r^{(n+1)}
	\le
	\widehat q_{r,1-\alpha_r}
	\right\}.
	\]
	We next show that $\mathcal E$ is contained in the joint position coverage
	event in \eqref{eq:joint_position_coverage_requirement} by considering the
	following two cases.
	
	\emph{Case 1:}
	$\widehat q_{\theta,1-\alpha_\theta}<+\infty$
	and
	$\widehat q_{r,1-\alpha_r}<+\infty$.
	On the event $\mathcal E$, both
	$S_\theta^{(n+1)}$ and $S_r^{(n+1)}$ are finite.
	Hence, by \eqref{eq:theta_score_matched} and
	\eqref{eq:range_score_matched},
	$\widehat K^{(n+1)}\ge K^{(n+1)}$, and the deterministic injective
	matching
	$\pi_{n+1}^\star\in
	\mathfrak I(K^{(n+1)},\widehat K^{(n+1)})$
	is well defined. Moreover, for every
	$k\in[K^{(n+1)}]$, the event $\mathcal E$ implies
	\begin{align}
		\left|
		\theta_{k0}^{(n+1)}
		-
		\widehat\theta_{\pi_{n+1}^\star(k),0}^{(n+1)}
		\right|
		&\le
		\widehat q_{\theta,1-\alpha_\theta},
		\label{eq:theta_matched_event}\\
		\left|
		r_{k0}^{(n+1)}
		-
		\widehat r_{\pi_{n+1}^\star(k),0}^{(n+1)}
		\right|
		&\le
		\widehat q_{r,1-\alpha_r}.
		\label{eq:range_matched_event}
	\end{align}
	It follows from
	\eqref{eq:CP_theta_source_set_construction}--\eqref{eq:CP_cartesian_set_construction}
	that
	\[
	\boldsymbol p_k^{(n+1)}
	\in
	\mathcal C_{p,\pi_{n+1}^\star(k)}(\widehat{\boldsymbol R}^{(n+1)}),
	\qquad
	\forall k\in[K^{(n+1)}].
	\]
	In this case, $\widehat K_{\mathrm{final}}^{(n+1)}=\widehat K^{(n+1)}$, so the joint position coverage event holds on $\mathcal E$.
	
	\emph{Case 2:}
	$\widehat q_{\theta,1-\alpha_\theta}=+\infty$
	or
	$\widehat q_{r,1-\alpha_r}=+\infty$.
	According to \eqref{eq:CP_output_collection}, the final source-number estimate is $\widehat K_{\mathrm{final}}^{(n+1)}=K_{\max}$, and the method returns $K_{\max}$ indexed predicted regions, each equal to $\mathcal P_{\mathrm{all}}$.
	Since $K^{(n+1)}\le K_{\max}$,
	$\mathfrak I(K^{(n+1)},K_{\max})\neq\varnothing$.
	Furthermore, by \eqref{eq:entire_position_region},
	$\boldsymbol p_k^{(n+1)}\in\mathcal P_{\mathrm{all}}$
	for all $k\in[K^{(n+1)}]$.
	Thus, for any
	$\pi\in\mathfrak I(K^{(n+1)},K_{\max})$,
	\[
	\boldsymbol p_k^{(n+1)}
	\in
	\mathcal C_{p,\pi(k)}(\widehat{\boldsymbol R}^{(n+1)}),
	\qquad
	\forall k\in[K^{(n+1)}].
	\]
	Hence, the joint position coverage event holds  in this case.
	
	Therefore, in both cases,
	\begin{align}
		\mathcal E
		\subseteq
		&	\Bigl\{
		\exists\pi\in
		\mathfrak I\bigl(
		K^{(n+1)},\widehat K_{\mathrm{final}}^{(n+1)}
		\bigr):
		\notag\\
		&\quad
		\boldsymbol p_k^{(n+1)}
		\in \mathcal C_{p,\pi(k)}(\widehat{\boldsymbol R}^{(n+1)}),
		\ \forall k\in[K^{(n+1)}]
		\Bigr\}.
		\label{eq:score_event_position_event_inclusion}
	\end{align}
	
	Combining \eqref{eq:joint_score_coverage} and
	\eqref{eq:score_event_position_event_inclusion}, we obtain
	\begin{align}
		\mathbb P\Bigl(
		\boldsymbol p_k^{(n+1)}
		\in \mathcal C_{p,\pi(k)}(\widehat{\boldsymbol R}^{(n+1)}),
		\forall k\in[K^{(n+1)}]
		\Bigr)
		\ge 1-\alpha_p,
	\end{align} 
	which proves \eqref{eq:joint_position_coverage_requirement}.

	\bibliographystyle{IEEEtran}
	\bibliography{ref}
	
\end{document}